\documentclass[a4paper,11pt]{article}
\usepackage{jcappub}
\usepackage{lineno}
\usepackage{bm}
\usepackage{xcolor}

\newcommand{\LCDM}{$\Lambda$CDM}
\newcommand{\wowa}{$w_0w_a$CDM}
\newcommand{\Lc}{\Lambda_c}
\newcommand{\ac}{a_c}
\newcommand{\vc}{v_c}
\newcommand{\fbde}{f_{\rm BDE}}
\newcommand{\sig}{\sigma_8}
\newcommand{\Seight}{S_8}
\newcommand{\Hub}{H_0}
\newcommand{\Next}{N_{\rm ext}}
\newcommand{\khalf}{k_{1/2}}

\usepackage{amsmath}
\usepackage[section]{placeins}          

\title{The unified bound dark energy--bound dark matter model confronted with DESI DR2, ACT and SPT-3G, and
the Unite supernovae measurements}

\author[a,1]{Jose Agustin Lozano Torres\note{Corresponding author.}}
\author[a]{and Axel de la Macorra}
\affiliation[a]{Instituto de F\'isica, Universidad Nacional Aut\'onoma de M\'exico,\\
Ciudad de M\'exico, 04510, M\'exico}
\emailAdd{lozano@estudiantes.fisica.unam.mx}

\abstract{The unified bound dark energy--bound dark matter (BDE-BDM) model derives the whole dark sector from a
single supersymmetric SU(3) dark gauge group, whose lightest meson is the dark energy and whose heavier bound
states, created with a velocity $\vc$ when the group condenses, are the dark matter. Because the condensation scale
$\Lc$ is fixed by gauge-coupling unification, the dark-energy sector has no free parameter beyond those of \LCDM,
and the dark-matter sector adds only $\vc$. We confront the model with DESI DR2 baryon acoustic oscillations (BAO),
the \textit{Planck}, ACT DR6 and SPT-3G D1 CMB spectra, the combined ACT and SPT CMB lensing and the Unite
supernovae. The measured condensation scale, $\Lc=43.74\pm0.11$~eV, lies inside the window $34^{+16}_{-11}$~eV
predicted by unification, and the resulting dark energy thaws without ever crossing the phantom divide, reaching
$w_0=-0.9267\pm0.0003$ today; it thus reproduces the DESI distances that \wowa{} fits with a phantom past. BDE-BDM
gives the best fit to the DESI DR2 BAO of the four models considered, reducing $\chi^2_{\rm BAO}$ by $20\%$ with
respect to \LCDM, more than \wowa, and over the full data combination it is statistically equivalent to both
\LCDM{} and \wowa, with $\ln B=-0.50$ and $-0.15$, while having two parameters fewer than \wowa{} in the
dark-energy sector. Within bound dark energy, bound dark matter is moderately preferred over cold dark matter,
$\ln B=+3.2$ with $\Delta\chi^2=-9.0$, a preference that comes mostly from the CMB. The data now constrain the bound
velocity from both sides, $\vc=0.37\pm0.14$, which places the free-streaming cutoff of the matter power spectrum at
$\khalf=1.49\,h\,{\rm Mpc}^{-1}$ and lowers the amplitude of structure to $\Seight=0.807\pm0.007$.
Future high-redshift measurements and small-scale data such as the Lyman-$\alpha$ forest, together with growth and lensing experiments, will therefore provide the decisive tests of the unified dark sector.}


\begin{document}
\maketitle
\flushbottom

\section{Introduction}
\label{sec:intro}

For over two decades, the accelerated expansion of the Universe has been attributed to a cosmological constant. The
resulting \LCDM{} model agrees with a broad range of observations, but the constant has no established physical
origin, and its measured value lies far below the expectations of quantum field theory. The second data release of
the Dark Energy Spectroscopic Instrument (DESI) measures the baryon acoustic oscillation (BAO) scale with sub-percent
precision over $0.1<z<4.2$~\cite{DESI2025BAO,DESI2025II}; combined with the cosmic microwave background (CMB) and
type Ia supernovae, these measurements prefer a dark energy whose equation of state evolves with time, at a
significance that depends on the supernova sample~\cite{DESI2025II}. The usual description of this preference, the
\wowa{} parametrization $w(a)=w_0+w_a(1-a)$, fits the data with $w_0>-1$ and $w_a<0$, an equation of state that
crosses $w=-1$ at low redshift; it is, however, a fitting function, whose parameters are free and not derived from
any underlying theory.

The CMB side of this comparison has also changed. The ACT DR6~\cite{Louis2025} and SPT-3G D1~\cite{Camphuis2025}
temperature and polarization spectra extend the \textit{Planck} measurements~\cite{Planck2020V,Rosenberg2022} to small
angular scales, and the combined ACT and SPT lensing reconstruction~\cite{ACTSPTlens2025} measures the amplitude of
structure. The Unite compilation~\cite{Camilleri2026,Lee2026} combines the Pantheon+ and DES 5-year supernova samples in a single
calibration.

In this paper we confront these data with the unified bound dark energy--bound dark matter (BDE-BDM)
model~\cite{LozanoMacorra2026PRD}, in which the dark energy and the whole of the dark matter originate from a single
supersymmetric SU($N_c$) gauge group that remains decoupled from the Standard Model. The dynamics are analogous to
those of quantum chromodynamics: as the Universe cools, the coupling of the dark gauge group grows until the
elementary fields confine into neutral composite states, as quarks and gluons form hadrons. The lightest dark
mesons behave as a canonical scalar field whose potential, generated by the gauge dynamics, drives the late-time
acceleration; this is bound dark energy (BDE)~\cite{Macorra2018,Almaraz2019,MacorraLozano2025,MacorraLozano2026JCAP}.
The heavier composite states acquire their mass from the binding and form the dark matter, bound dark matter
(BDM)~\cite{Macorra2010,Mastache2020}; created relativistically, they stream out of small-scale perturbations and suppress
the matter power spectrum below a scale set by their velocity at formation, $\vc$. A single condensation therefore
accounts for the late-time acceleration and predicts the small-scale behavior of the dark matter.

The model's theoretical framework relies on the condensation scale. Gauge-coupling unification gives
$\Lc=34^{+16}_{-11}$~eV, and entropy conservation fixes the scale factor of the condensation, $\ac$, once $\Lc$ is
known. Once $\Lc$ is specified, the dark-energy sector has no further freedom: the present dark-energy density, the
evolution of its equation of state and the epoch of the transition are derived aspects of the model. The dark-matter sector
adds one parameter, $\vc$, which sets the free-streaming suppression; unlike in warm dark matter scenarios, the epoch
at which the dark matter becomes non-relativistic is tied to the same condensation that produces the dark energy. In ref.~\cite{LozanoMacorra2026PRD} we presented the full theory of the unified model, including its linear
perturbations and the energy-conserving division of the condensation energy between the two bound components, where in there we constrained it with DESI DR2, \textit{Planck} and three older supernova compilations. In this work we test whether the unified model remains viable when confronted with the latest CMB data and the
Unite supernova compilation, we isolate the effect of the bound dark matter by comparing BDE-BDM with BDE-CDM fitted
to identical data, and we determine how strongly the new small-scale CMB and lensing measurements constrain the
velocity of the bound states at their formation. We find that BDE-BDM gives the best fit to the DESI DR2 BAO of the
models considered and is statistically equivalent to \LCDM{} and \wowa{} over the full data combination, that bound
dark matter is moderately preferred over cold dark matter, and that the bound velocity is now constrained from both
sides.

The paper is organized as follows. Section~\ref{sec:model} presents the theoretical framework of the unified model,
section~\ref{sec:methods} the datasets and methodology, section~\ref{sec:results} the results and observational
constraints, section~\ref{sec:comparison} the model comparison, and section~\ref{sec:conclusions} our conclusions.

\section{Theoretical framework of the unified BDE-BDM model}
\label{sec:model}

We summarize the model; the derivations, the linear perturbations, and the numerical implementation are given in
ref.~\cite{LozanoMacorra2026PRD}. As in the BDE and BDM analyses, the subscript $c$ denotes quantities evaluated at
the condensation scale factor $\ac$; the dark plasma is described for $a\le\ac$ and the two bound components for
$a\ge\ac$.

\subsection{The dark gauge sector}

The dark sector is a supersymmetric SU($N_c$) gauge theory with $N_c=3$ colors and $N_f=6$ flavors of massless dark
quarks, interacting with the Standard Model only gravitationally~\cite{Macorra2018,Almaraz2019}. Above the
condensation scale, the theory is asymptotically free, and its elementary particles, dark gluons, dark quarks and
their superpartners, form a relativistic dark plasma with
\begin{equation}
  g_{\rm DG}=\left(1+\frac{7}{8}\right)\left[2(N_c^2-1)+2N_cN_f\right]=97.5
  \label{eq:gDG}
\end{equation}
effective degrees of freedom. Unifying its coupling with the gauge couplings of the minimal supersymmetric Standard
Model at $\Lambda_{\rm GUT}=(1.05\pm0.07)\times10^{16}$~GeV, the coupling becomes strong at
\begin{equation}
  \Lc=\Lambda_{\rm GUT}\,e^{-8\pi^2/(b_0 g_{\rm GUT}^2)}=34^{+16}_{-11}~{\rm eV},
  \qquad b_0=3N_c-N_f=3,
  \label{eq:Lc}
\end{equation}
The uncertainty is dominated by that of the unification coupling through the exponential, as the QCD scale
follows from the running of the strong coupling. The dark sector decouples from the Standard Model at unification,
and entropy conservation in each sector, with the plasma density matched at condensation to
$\rho_{\rm DG}(\ac)=3\Lc^4$, relates the condensation scale factor to $\Lc$~\cite{Almaraz2019,LozanoMacorra2026PRD},
\begin{equation}
  \ac\Lc=1.0939\times10^{-4}~{\rm eV}.
  \label{eq:acLc}
\end{equation}
The normalization fixes the plasma temperature at condensation,
$T_c=[90/(\pi^2g_{\rm DG})]^{1/4}\Lc\simeq0.55\,\Lc$, slightly below the scale at which the coupling diverges. Before
condensation, the dark plasma adds to the expansion rate the equivalent of $\Next=0.945$ effective neutrino species.
This contribution is fixed by the group content rather than fitted, and it is transient: after $\ac$ the extra
energy from the dark plasma is converted into the two bound components, whose combined density dilutes faster than radiation, so that
the radiation content around recombination is essentially standard.

\subsection{Bound dark energy}

At the transition, the lightest bound state is a dark meson, described by a canonically normalized scalar field
$\phi$ whose potential is generated non-perturbatively by the gauge dynamics and takes the inverse-power-law form
\begin{equation}
  V(\phi)=\Lc^{4+n}\phi^{-n},\qquad n=2\left[1+\frac{2}{N_c-N_f}\right]=\frac{2}{3},
  \label{eq:Vphi}
\end{equation}
where both the exponent and the overall scale follow from the group content and eq.~\eqref{eq:Lc}. The associated
mass, $m_\phi^2=(10/9)(\Lc/\phi)^{8/3}\Lc^2$, is protected by supersymmetry and by the global symmetries of the
condensate~\cite{Almaraz2019}. No parameter of the dark-energy sector is chosen by hand, which distinguishes the model
from quintessence with the same form of $V(\phi)$, where the scale of the potential is adjusted to reproduce the
observed acceleration.

The field starts at $\ac$ with the energy inherited from the fraction $\fbde$ of the plasma that condenses into the
meson, the remainder forming the bound dark matter:
\begin{equation}
  \phi_c=\Lc,\qquad V(\phi_c)=\Lc^4,\qquad \rho_{\rm BDE}(\ac)=\fbde\,\rho_{\rm DG}(\ac)=3\fbde\Lc^4.
  \label{eq:initial}
\end{equation}
Since $V(\phi_c)$ is fixed by the symmetry-breaking scale, the partition determines the kinetic term,
$\dot\phi_c^2/2=(3\fbde-1)\Lc^4$, and with it the equation of state at the transition,
\begin{equation}
  w_{\rm BDE}(\ac)=\frac{3\fbde-2}{3\fbde},\qquad \frac{1}{3}\le\fbde\le1,
  \label{eq:wac}
\end{equation}
where the lower limit, positivity of the kinetic term, is a physical bound rather than a prior: the meson cannot
inherit less than $\Lc^4$, so the bound dark matter cannot take more than two thirds of the condensation energy. In
the limit $\fbde\to1$ the initial conditions reduce to those of BDE-CDM~\cite{Almaraz2019,MacorraLozano2026JCAP}.
Thereafter the field obeys the Klein--Gordon equation in the expanding background, and its history has three
stages. Immediately after $\ac$ the kinetic term dominates and the field enters a stiff phase, $w\simeq1$, in which
its density dilutes as $a^{-6}$ and becomes negligible within a few $e$-folds. The field then freezes by Hubble
friction, with $w\simeq-1$ and a density far below the background. Finally, when the background density falls to the
level of the frozen field, at redshifts of order unity, the potential takes over and the field thaws: $w$ rises
from $-1$ toward a present value that stays above the phantom divide. This late-time history is a function of $\Lc$
alone, and the expansion it generates is of the thawing, non-phantom kind preferred by the DESI data.

\subsection{Bound dark matter}

The heavier bound states acquire their mass from the binding, as most of the proton mass arises from QCD rather than
from the quark masses~\cite{Macorra2010,Mastache2020}. In the unified model they inherit the complement of the
condensing fraction,
\begin{equation}
  \begin{aligned}
  \rho_{\rm BDM}(\ac)&=3(1-\fbde)\Lc^4,\\
  \lambda_{\rm fs}\propto\left[3(1-\fbde)\right]^{-1/3}\Lc^{-4/3},&\qquad
  M_{\rm fs}\propto\left[3(1-\fbde)\right]^{-1}\Lc^{-4},
  \end{aligned}
  \label{eq:rhobdm}
\end{equation}
so that the free-streaming length and mass of the dark matter and the late-time acceleration are controlled by the
same $\Lc$, together with the partition at condensation; as an equivalent energy scale,
$\rho_{\rm BDM}(\ac)^{1/4}\simeq0.66\,\Lc$ for the parameters preferred by the data. Each particle emerges from the
transition with velocity $\vc$, and its momentum then redshifts freely, $p\propto1/a$, so that
\begin{equation}
  \begin{aligned}
  v(a)&=\frac{x}{\sqrt{1+x^2}},\qquad x\equiv\gamma_c\vc\frac{\ac}{a},\qquad \gamma_c=\frac{1}{\sqrt{1-\vc^2}},\\
  w_{\rm BDM}(a)&=\frac{v^2}{3}=\frac{1}{3}\left[1+\frac{1-\vc^2}{\vc^2}\left(\frac{a}{\ac}\right)^2\right]^{-1}
  \qquad(a\ge\ac),
  \end{aligned}
  \label{eq:wbdm}
\end{equation}
with $w_{\rm BDM}(\ac)=\vc^2/3$, while $w_{\rm BDM}=1/3$ for $a<\ac$; today $w_{\rm BDM}\sim10^{-12}$. Energy
conservation then integrates in closed form,
\begin{equation}
  \rho_{\rm BDM}(a)=\rho_{\rm BDM,0}\,a^{-3}\,\frac{\gamma(a)}{\gamma(1)},\qquad
  \gamma(a)\equiv\sqrt{1+\gamma_c^2\vc^2\left(\frac{\ac}{a}\right)^2},
  \label{eq:rhobdma}
\end{equation}
which tends to pressureless matter well after the transition, and the adiabatic sound speed is
$c_a^2=w(5-3w)/[3(1+w)]$, interpolating between $1/3$ in the relativistic limit and zero in the cold limit. For
$\vc\to0$ the component is cold at the transition and indistinguishable from cold dark matter on the scales resolved
here; for $\vc$ approaching $1/\sqrt2$, the value at which the momentum equals the mass, it stays semi-relativistic
for several $e$-folds and behaves as a warm species. The velocity $\vc$ is the single free parameter of the
dark-matter sector.

\subsection{The partition of the condensation energy}

The plasma energy at the transition is shared between the two bound components without loss,
$\rho_{\rm DG}(\ac)=\rho_{\rm BDE}(\ac)+\rho_{\rm BDM}(\ac)$, which defines the condensing fraction
\begin{equation}
  \fbde=1-\frac{\rho_{\rm BDM}(\ac)}{\rho_{\rm DG}(\ac)}=1-\frac{\gamma_c\,\rho_{\rm BDM,0}\,\ac^{-3}}{3\Lc^4},
  \label{eq:fbde}
\end{equation}
where the second equality uses eq.~\eqref{eq:rhobdma}. Given the present dark-matter density $\omega_{\rm bdm}$ and
the velocity $\vc$, the fraction $\fbde$ is determined and is not an independent parameter. Through
eq.~\eqref{eq:wac} it fixes the initial equation of state of the dark energy, so that the dark-matter density and
velocity feed back on the dark-energy history; this coupling is what distinguishes BDE-BDM from BDE-CDM at fixed
$\Lc$. The model parameters are therefore those of \LCDM, with $\Lc$ in place of the cosmological constant and
$\omega_{\rm bdm}$ in place of $\omega_c$, plus $\vc$. The perturbations of the bound dark matter are evolved as an
imperfect fluid in the synchronous gauge~\cite{MaBertschinger1995}, with the matching at the transition derived in
ref.~\cite{LozanoMacorra2026PRD}.

\section{Datasets and methodology}
\label{sec:methods}

\paragraph{DESI DR2 BAO.} We use the baryon acoustic oscillation measurements of the second data release of
DESI~\cite{DESI2025II,DESI2025BAO}: thirteen distance ratios from seven tracer samples spanning $0.1<z<4.2$. The bright galaxy
sample at $z_{\rm eff}=0.295$ provides the isotropic distance $D_V/r_{\rm d}$; the luminous red galaxies at
$z_{\rm eff}=0.510$ and $0.706$, the combined luminous red and emission-line galaxies at $0.934$, the emission-line
galaxies at $1.321$, the quasars at $1.484$ and the Lyman-$\alpha$ forest at $2.330$ each provide the transverse and
radial distances $D_M/r_{\rm d}$ and $D_H/r_{\rm d}$, with their correlations. These measurements fix the expansion
history relative to the sound horizon $r_{\rm d}$, and are the data most directly sensitive to the evolution of the
dark energy.

\paragraph{CMB2026.} We combine three CMB experiments that together cover the temperature and polarization
anisotropies from the largest scales to the damping tail. From \textit{Planck} we use the PR4 CamSpec high-$\ell$
TT, TE and EE likelihood~\cite{Rosenberg2022}, the low-$\ell$ temperature likelihood~\cite{Planck2020V} and the
Sroll2 low-$\ell$ EE likelihood~\cite{Pagano2020}, which constrains the optical depth to reionization. The ACT DR6
CMB-only likelihood~\cite{Louis2025} extends the temperature and polarization spectra to smaller angular scales, and
the SPT-3G D1 likelihood~\cite{Camphuis2025} adds TT spectra over $425\lesssim\ell\lesssim2975$ and TE and EE spectra
over $425\lesssim\ell\lesssim3975$ from the SPT-3G Main field. We refer to this combination as CMB2026. The small
angular scales probed by ACT and SPT-3G are sensitive to the composition and expansion rate of the Universe before
recombination and to the amplitude of the primordial spectrum.

\paragraph{CMB lensing.} The combined ACT DR6 and SPT-3G lensing likelihood~\cite{ACTSPTlens2025} measures the
lensing potential power spectrum, an integrated probe of the amplitude of structure at intermediate redshift and
therefore sensitive to the suppression of power by the bound dark matter.

\paragraph{Unite supernovae.} The Unite compilation~\cite{Camilleri2026,Lee2026} combines the Pantheon+ and DES 5-year samples of
type Ia supernovae~\cite{Brout2022,DESSN5YR2024,DESDovekie} in a single calibration; we use its 2884 supernovae with
$z_{\rm HD}>0$. Uncalibrated supernovae
constrain the shape of the luminosity distance, and hence the late-time evolution of the dark energy, but not the
absolute scale $\Hub$. The Unite distances are bias-corrected and are not anchored to an absolute distance scale; in
accordance with the conditions of their release, we do not use them to infer $\Hub$, and all statements on $\Hub$ in
this paper are based on the combination without supernovae (section~\ref{sec:h0}). We refer to the full combination as DESI DR2 BAO $+$ CMB $+$ Unite.

Theory predictions are computed with CAMB~\cite{Lewis2000} and modified versions implementing BDE and BDE-BDM, with
recombination computed with CosmoRec~\cite{ChlubaThomas2011}, and the parameter space is explored with
Cobaya~\cite{Torrado2021}. The sampled parameters and their priors are listed in table~\ref{tab:priors}. \LCDM{} and
\wowa{} sample the angular size of the sound horizon, $100\theta_{\rm MC}$; in the BDE models $\Lc$ takes its place,
and $\Hub$ is derived. The dark-matter density has the same prior in all models, so that no prior-width difference
enters the Bayes factors. For $\vc$ we use $\mathcal{U}[10^{-3},0.99]$; its posterior lies entirely below $1/\sqrt2$,
so the upper limit does not affect any parameter constraint. All models include one massive neutrino with
$\sum m_\nu=0.06$~eV and $N_{\rm eff}=3.044$. For BDE-BDM we also analyze DESI DR2 BAO $+$ CMB2026 $+$ CMB lensing without supernovae, which we use for the
comparison of $\Hub$ with the local distance ladder and to test the robustness of the constraints on $\vc$. Chains
are run until the Gelman--Rubin statistic satisfies $R-1<0.01$,
and best fits are obtained with several minimizations from different starting points, and the posteriors are analyzed with GETDIST~\cite{Lewis_2025}.

\begin{table}[!htb]
\centering
\small
\begin{tabular}{l l l}
\hline
Parameter & Prior & Models \\
\hline
$\omega_b$ & $\mathcal{U}[0.005,\,0.1]$ & all \\
$\omega_c$ ($\omega_{\rm bdm}$ in BDE-BDM) & $\mathcal{U}[0.001,\,0.99]$ & all \\
$100\theta_{\rm MC}$ & $\mathcal{U}[0.5,\,10]$ & \LCDM, \wowa \\
$\Lc$ [eV] & $\mathcal{U}[23,\,50]$ & BDE-CDM, BDE-BDM \\
$\vc$ & $\mathcal{U}[0.001,\,0.99]$ & BDE-BDM \\
$\tau_{\rm reio}$ & $\mathcal{U}[0.01,\,0.8]$ & all \\
$n_s$ & $\mathcal{U}[0.8,\,1.2]$ & all \\
$\ln(10^{10}A_s)$ & $\mathcal{U}[1.61,\,3.91]$ & all \\
$w_0$ & $\mathcal{U}[-3,\,1]$ & \wowa \\
$w_a$ & $\mathcal{U}[-3,\,2]$ & \wowa \\
\hline
\end{tabular}
\caption{\label{tab:priors}%
Sampled parameters and their uniform priors.}
\end{table}

We compare models through the best-fit $\chi^2$, the deviance information criterion,
${\rm DIC}=\overline{D}+p_D$ with $p_D=\overline{D}-D(\hat\theta)$~\cite{Spiegelhalter2002}, and the Bayesian evidence
computed with MCEvidence~\cite{Heavens2017} from the same chains. As in ref.~\cite{Lodha2025}, the evidences of the
models that sample $\theta_{\rm MC}$ are referred to $\Hub\sim\mathcal{U}[20,100]$ through the Jacobian of the
transformation, so that the geometric degree of freedom carries the same prior in all models; we quote $\ln B$ on the
scale of ref.~\cite{Trotta2008}.

As a check of our implementation of the data combination, our \LCDM{} and \wowa{} constraints reproduce those of the
Unite analysis for Unite $+$ BAO $+$ CMB~\cite{Camilleri2026}, obtained with a different sampler and slightly different
CMB and lensing likelihoods: $\Omega_m=0.3039\pm0.0032$ against $0.303\pm0.003$ for \LCDM, and
$(w_0,w_a)=(-0.864\pm0.043,\,-0.60\pm0.18)$ against $(-0.861^{+0.044}_{-0.042},\,-0.60^{+0.17}_{-0.19})$ for \wowa, with
an improvement of the fit over \LCDM{} of $\Delta\chi^2=-13.2$ here and $-12.4$ in ref.~\cite{Camilleri2026}.

\section{Results and observational constraints}
\label{sec:results}

\subsection{Internal consistency of the BDE-BDM model}
\label{sec:consistency}

\subsubsection{\texorpdfstring{BDE-BDM key parameters: $\Lc$, $\ac$, $\fbde$ and $\vc$}{BDE-BDM key parameters: Lambda\_c, a\_c, f\_BDE and v\_c}}
\label{sec:keyparams}

\begin{figure}[!tbp]
\centering
\includegraphics[width=0.8\textwidth]{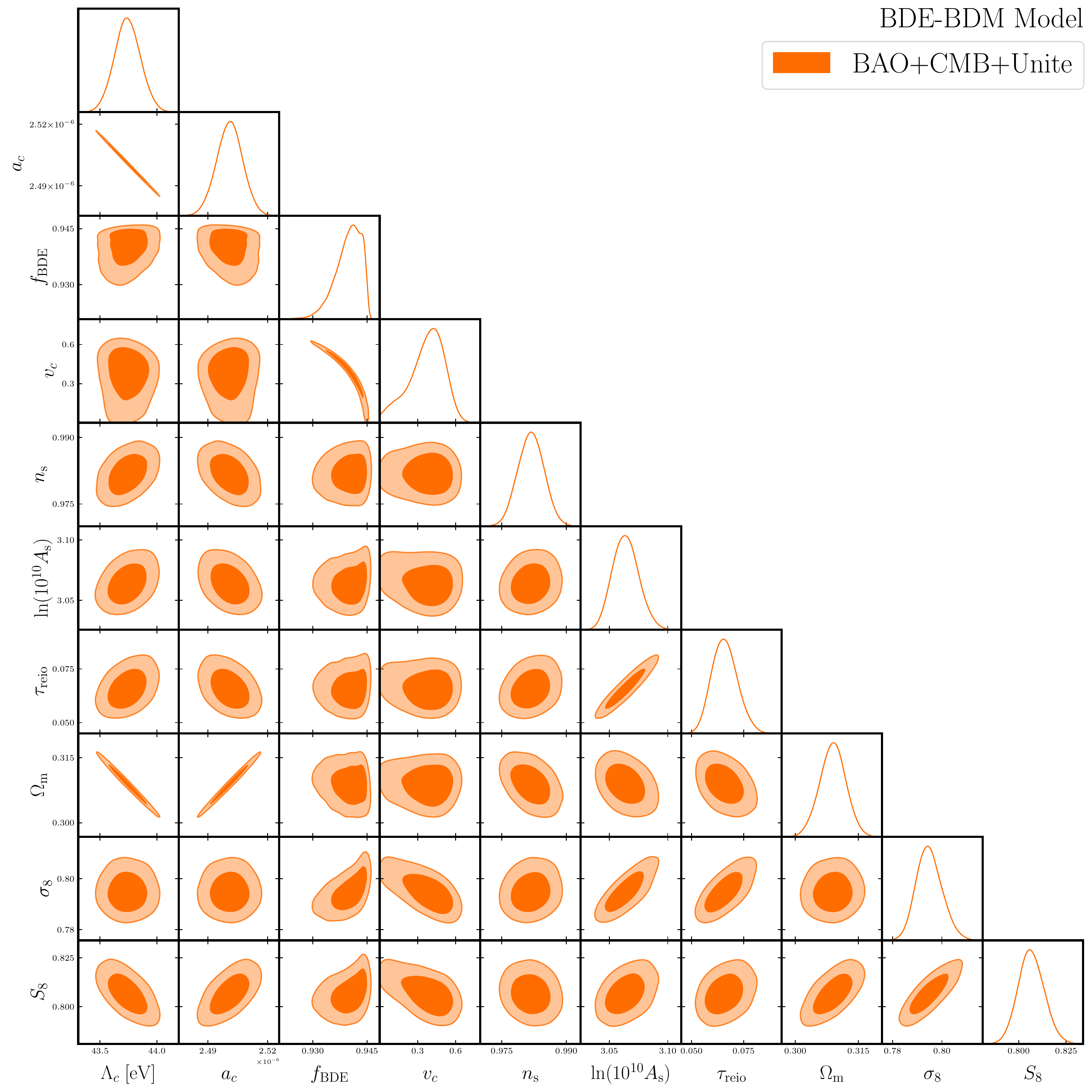}
\caption{\label{fig:tribdm}%
Marginalized $68\%$ and $95\%$ credible regions of the BDE-BDM parameters for DESI DR2 BAO $+$ CMB2026 $+$ Unite $+$
CMB lensing.}
\end{figure}

Figure~\ref{fig:tribdm} shows the joint posteriors of the BDE-BDM parameters. The condensation scale and the
condensation scale factor are
\begin{equation}
  \Lc=43.74\pm0.11~{\rm eV},\qquad \ac=(2.5009\pm0.0065)\times10^{-6},
  \label{eq:Lcfit}
\end{equation}
anticorrelated as required by eq.~\eqref{eq:acLc}. The condensation scale lies inside the window predicted by
eq.~\eqref{eq:Lc}, and agrees with the BDE-CDM value $\Lc=43.81\pm0.11$~eV obtained from the same data. The
condensing fraction is $\fbde=0.9398\pm0.0037$, well inside the physical range of eq.~\eqref{eq:wac}: the bound
dark matter takes $6\%$ of the condensation energy, and the dark energy emerges with
$w_{\rm BDE}(\ac)=(3\fbde-2)/(3\fbde)=0.29$, just below the plasma value $1/3$.

The bound velocity is now constrained from both sides,
\begin{equation}
  \vc=0.37\pm0.14,\qquad 0.066<\vc<0.60\quad(95\%),
  \label{eq:vcfit}
\end{equation}
whereas in our earlier analysis with DESI DR2 and \textit{Planck} its $95\%$ interval extended to the edge of the
prior~\cite{LozanoMacorra2026PRD}. The posterior has no support above $1/\sqrt2$, and the cold limit $\vc\to0$ is
excluded at the same $95\%$ level, consistent with the preference for bound over cold dark matter found in
section~\ref{sec:comparison}. The bound does not rely on the supernovae: without them, DESI DR2 BAO $+$ CMB2026 $+$
lensing give $\vc=0.370\pm0.142$, with $0.059<\vc<0.607$ at $95\%$, essentially the same interval, so the bound
velocity is constrained by the CMB and BAO. The degeneracy between $\vc$ and $\fbde$ visible in figure~\ref{fig:tribdm} follows
from eq.~\eqref{eq:fbde}: at fixed $\omega_{\rm bdm}$, a larger velocity raises $\gamma_c$ and hence the share of the
condensation energy carried by the bound dark matter.

The remaining correlations in figure~\ref{fig:tribdm} show how the dark-energy sector, controlled by the single scale
$\Lc$, is tied to the background. A higher $\Lc$ means an earlier condensation, eq.~\eqref{eq:acLc}, and a denser dark
energy today, so $\Lc$ is anticorrelated with $\Omega_m$, and the projected
$w_0$ and $w_a$ are functions of $\Lc$ and $\fbde$ rather than independent directions. The bound velocity $\vc$, in contrast, is only weakly correlated with the background quantities.

\subsubsection{Hubble constant and matter density}
\label{sec:h0}

In the BDE models $\Hub$ is not sampled: it follows from $\Lc$ through the background evolution. Because the
equation of state is predicted rather than fitted, no $\Hub$--$w$ degeneracy opens, unlike in \wowa. The matter density
is higher than in \LCDM, $\Omega_m=0.3089\pm0.0031$ against $0.3039\pm0.0032$.

 We do not report $\Hub$ for any combination that
includes the Unite supernovae, in the parameter table (table~\ref{tab:params}) or in the figures, for two reasons.
First, the Unite Hubble diagram is not anchored to an absolute distance scale. Type Ia supernovae without a
calibration constrain the shape of the luminosity distance, because their absolute magnitude is degenerate with
$\Hub$; in a joint fit with the BAO and the CMB, that degeneracy is broken only by the sound horizon, which the CMB
calibrates and the BAO transfer to low redshift. The supernova distances would then be placed on an absolute scale
through the sound horizon, which is the construction of an inverse distance ladder. Second, the Unite distances are
bias-corrected, including corrections for peculiar velocities and for the Eddington bias at very low redshift, and are
therefore not suitable for calibration against low-redshift distance anchors or for inferring $\Hub$. The supernovae
thus enter our analysis only through the shape of the distance--redshift relation, which constrains the matter density
and the evolution of the dark energy, and every statement on $\Hub$ in this work, including the comparison with the
local distance ladder below, is based on DESI DR2 BAO $+$ CMB2026 $+$ CMB lensing without supernovae, as in
ref.~\cite{LozanoMacorra2026MNRAS}. For the same reason, we show the BAO observables as the dimensionless distance ratios measured by DESI and do not
quote $H(z)$ in absolute units. This combination gives
\begin{equation}
  \Hub=67.22\pm0.22~{\rm km\,s^{-1}\,Mpc^{-1}},
  \label{eq:h0}
\end{equation}
with $r_{\rm d}=147.50\pm0.17$~Mpc. The value lies $5.5\sigma$ below the SH0ES measurement,
$\Hub=73.04\pm1.04$~km\,s$^{-1}$\,Mpc$^{-1}$~\cite{Riess2022}. The unified model therefore does not address the Hubble
tension: its dark energy is negligible before recombination and leaves the sound horizon unchanged, so the CMB and BAO
constrain $\Hub$ much as in \LCDM{} and the tension remains.

\subsubsection{\texorpdfstring{Equation of state in the BDE-BDM model: $w_0$ and $w_a$}{Equation of state in the BDE-BDM model: w0 and wa}}
\label{sec:w0wa}

\begin{figure}[!tbp]
\centering
\includegraphics[width=0.6\textwidth]{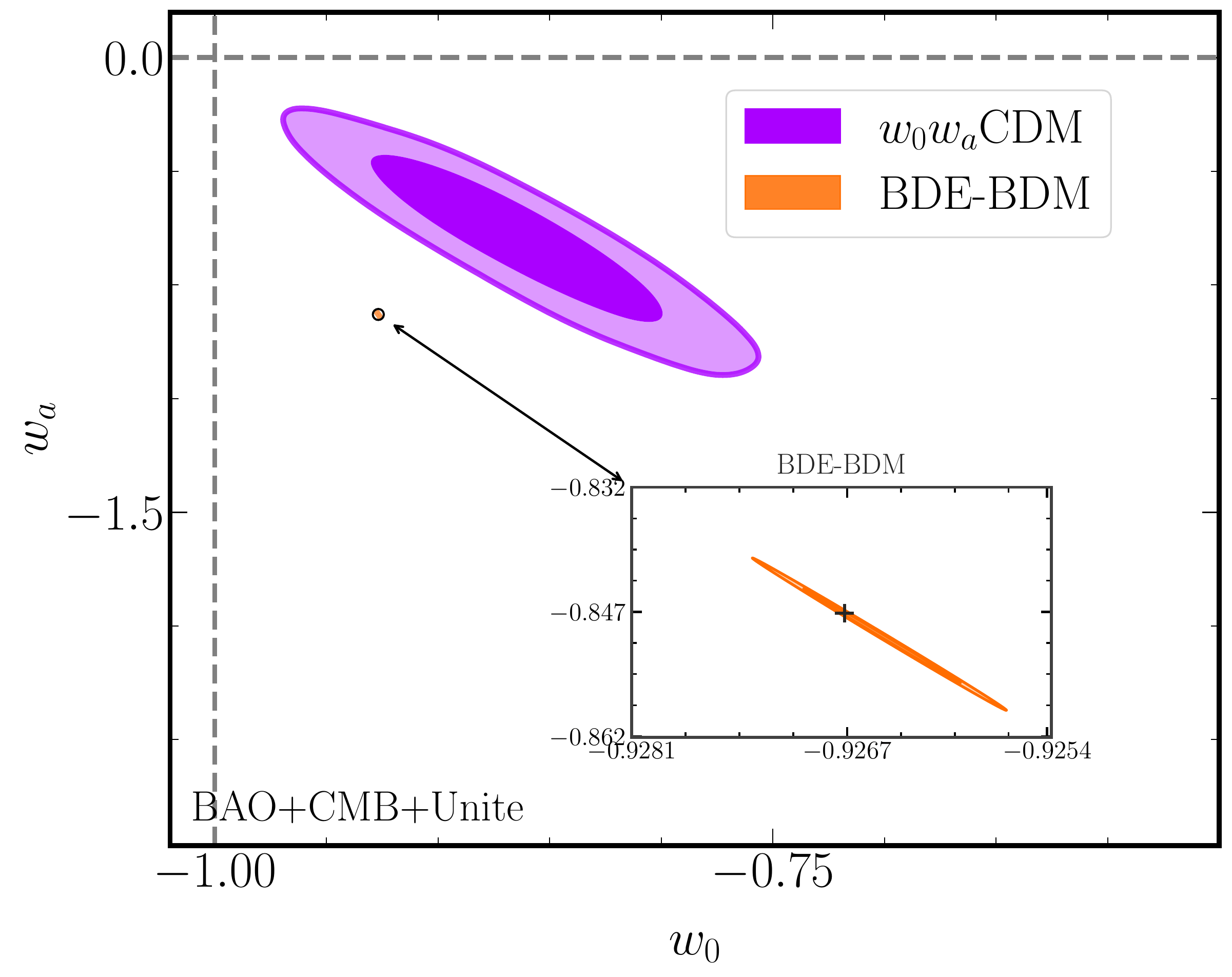}
\caption{\label{fig:w0wa}%
The $w_0$--$w_a$ plane: $68\%$ and $95\%$ regions of \wowa{} and the BDE-BDM projection of
eq.~\eqref{eq:w0waproj}, enlarged in the inset. The dashed lines mark \LCDM.}
\end{figure}

To compare with \wowa, we project the BDE-BDM equation of state onto its value and slope today~\cite{WolfGarcia2024},
\begin{equation}
  w_0\equiv w(a=1),\qquad w_a\equiv-\left.\frac{dw}{da}\right|_{a=1},
  \label{eq:w0waproj}
\end{equation}
which gives $(w_0,w_a)=(-0.9267\pm0.0003,\,-0.847\pm0.004)$ (figure~\ref{fig:w0wa}). These are derived quantities:
the constraint is a narrow segment because both follow from $\Lc$ and, through $\fbde$, from $\vc$. The point lies
outside the $95\%$ region of the \wowa{} fit, $w_0=-0.864\pm0.043$ and $w_a=-0.60\pm0.18$. This does not imply a
worse fit: the projection captures only the value and slope of $w$ today, and the BDE-BDM equation of state flattens
toward $-1$ at higher redshift instead of following the CPL form into the phantom regime, so the two models reach
similar distances through different histories (section~\ref{sec:eos}).

\subsubsection{Bound versus cold dark matter}
\label{sec:bdmcdm}

\begin{figure}[!tbp]
\centering
\includegraphics[width=0.54\textwidth]{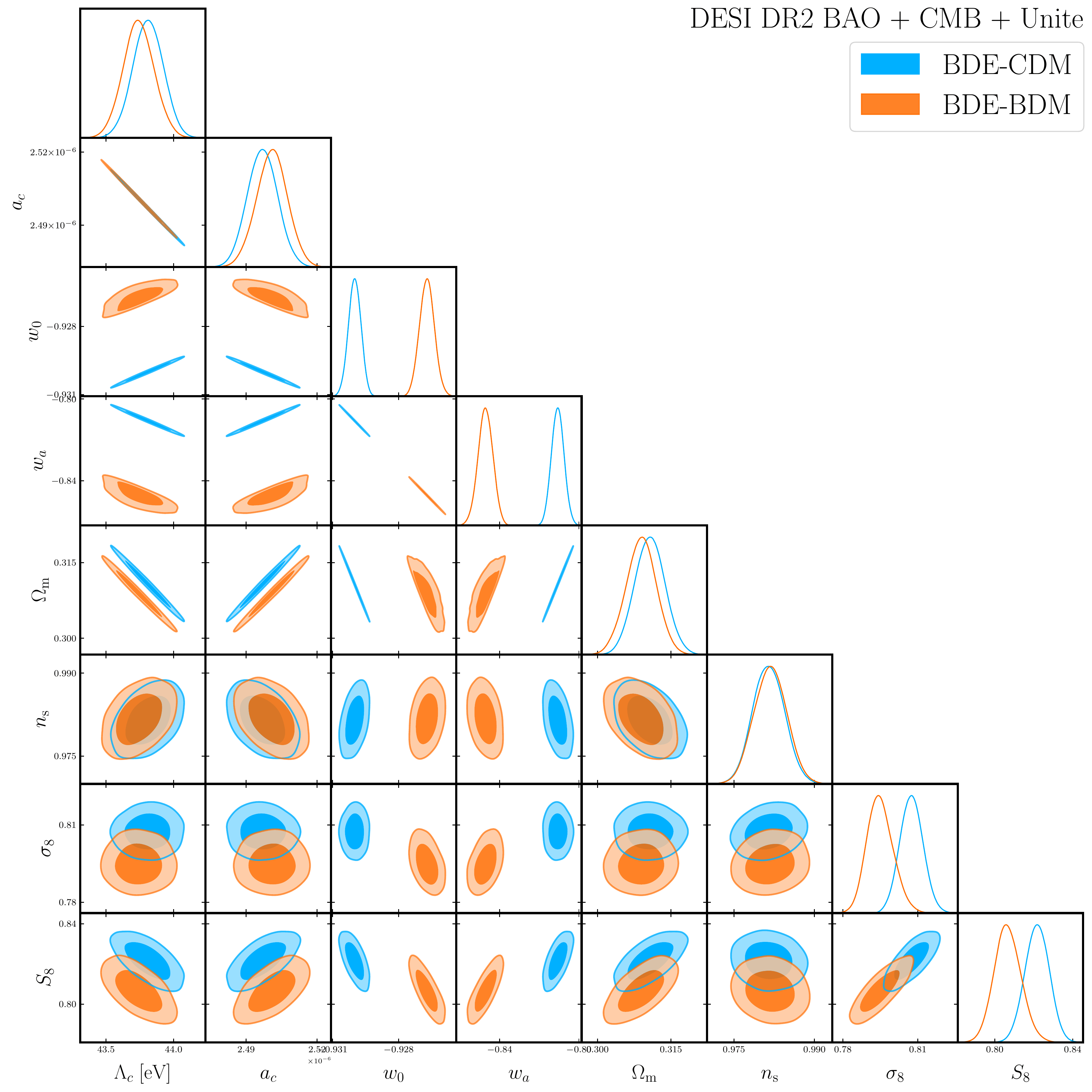}
\caption{\label{fig:bdecdm}%
BDE-CDM and BDE-BDM fitted to the same data, DESI DR2 BAO $+$ CMB2026 $+$ Unite $+$ CMB lensing: $68\%$ and $95\%$
credible regions.}
\end{figure}

Figure~\ref{fig:bdecdm} compares the two BDE models fitted to identical data, which isolates the effect of the bound
dark matter. The condensation scale, the expansion rate and the primordial parameters are essentially unchanged; the
bound dark matter lowers the amplitude of structure, $\sig=0.795\pm0.005$ and $\Seight=0.807\pm0.007$ against
$\sig=0.808\pm0.005$ and $\Seight=0.822\pm0.006$ in BDE-CDM, and shifts the projected $(w_0,w_a)$ slightly through the
energy partition of eq.~\eqref{eq:fbde}. As in ref.~\cite{LozanoMacorra2026PRD}, the lower amplitude has two origins: the free streaming of the bound dark
matter, which suppresses the power on small scales (section~\ref{sec:growth}), and the share of the condensation
energy it carries, which modifies the dark-energy history through eq.~\eqref{eq:wac}.

The two models agree on what the dark-energy sector fixes: the condensation scale differs by $0.4\sigma$,
$\Lc=43.81\pm0.11$~eV for BDE-CDM against $43.74\pm0.11$~eV, and so does the present matter density. None of the sampled parameters shifts by more than about $1\sigma$; the optical depth and the primordial
amplitude move by $-0.8\sigma$ and $-1.0\sigma$ toward their \LCDM{} values (section~\ref{sec:constraints}). The
larger differences are in derived quantities. The amplitude decreases by $0.013$ in $\sig$ and $0.015$ in $\Seight$
($1.8\sigma$ and $1.6\sigma$ of the combined width), and the projected equation of state shifts from
$(w_0,w_a)=(-0.9300\pm0.0003,\,-0.811\pm0.003)$ in BDE-CDM to $(-0.9267\pm0.0003,\,-0.847\pm0.004)$. These last shifts,
$+0.0033$ in $w_0$ and $-0.036$ in $w_a$, are many times the width of these tightly constrained derived quantities, but
small in absolute terms, and far smaller than the distance between either BDE model and the \wowa{} region of
figure~\ref{fig:w0wa}. The bound dark matter therefore does not act by displacing the background parameters: it adds
two physical effects, the free streaming and the partition of the condensation energy, which the data detect through
the CMB lensing and the high-multipole spectra (section~\ref{sec:comparison}).

\FloatBarrier
\subsection{\texorpdfstring{Evolution of the equation of state in \LCDM, \wowa, BDE-CDM and BDE-BDM}{Evolution of the equation of state in LCDM, w0waCDM, BDE-CDM and BDE-BDM}}
\label{sec:eos}

\begin{figure}[!tbp]
\centering
\includegraphics[width=0.495\textwidth]{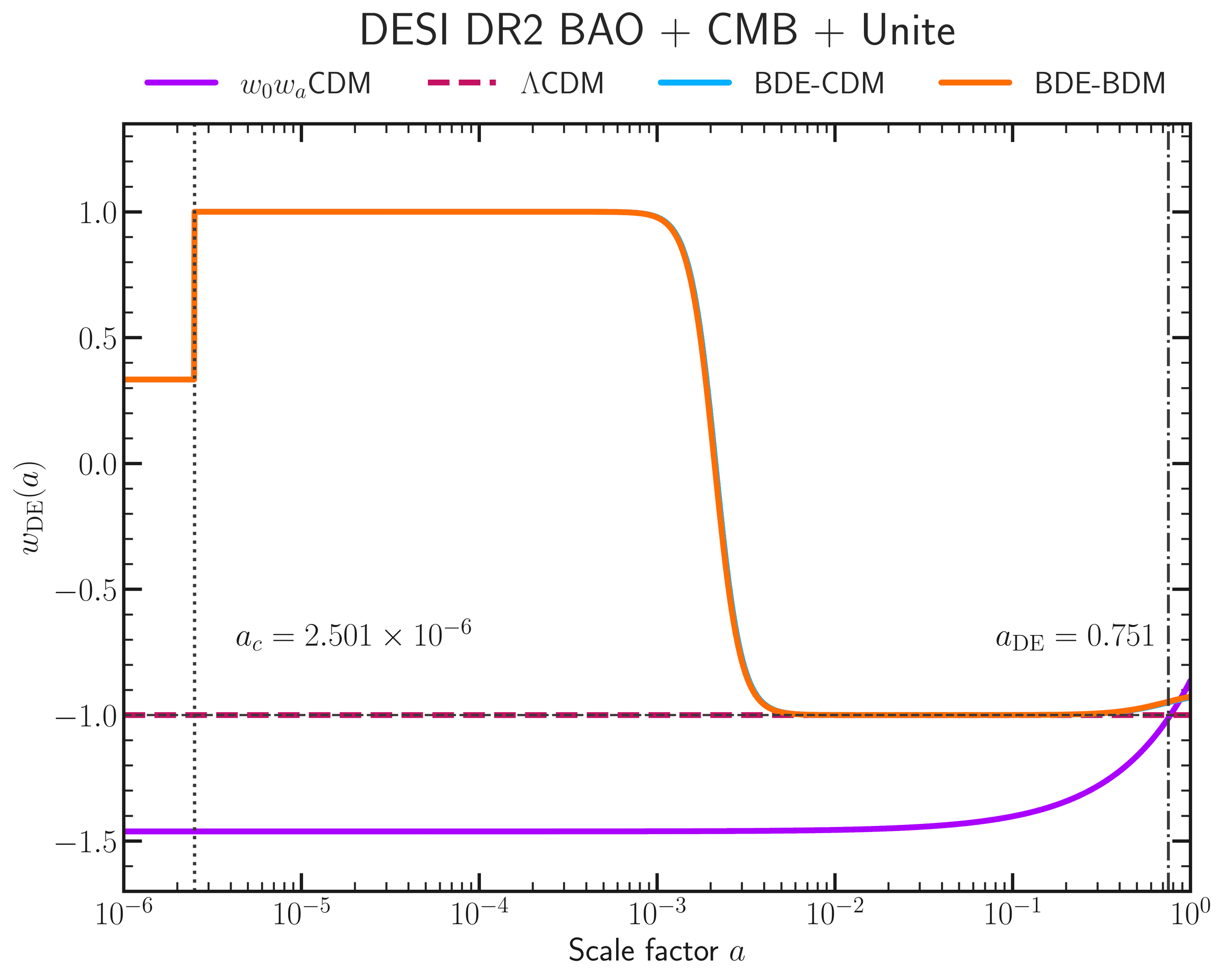}\hfill
\includegraphics[width=0.495\textwidth]{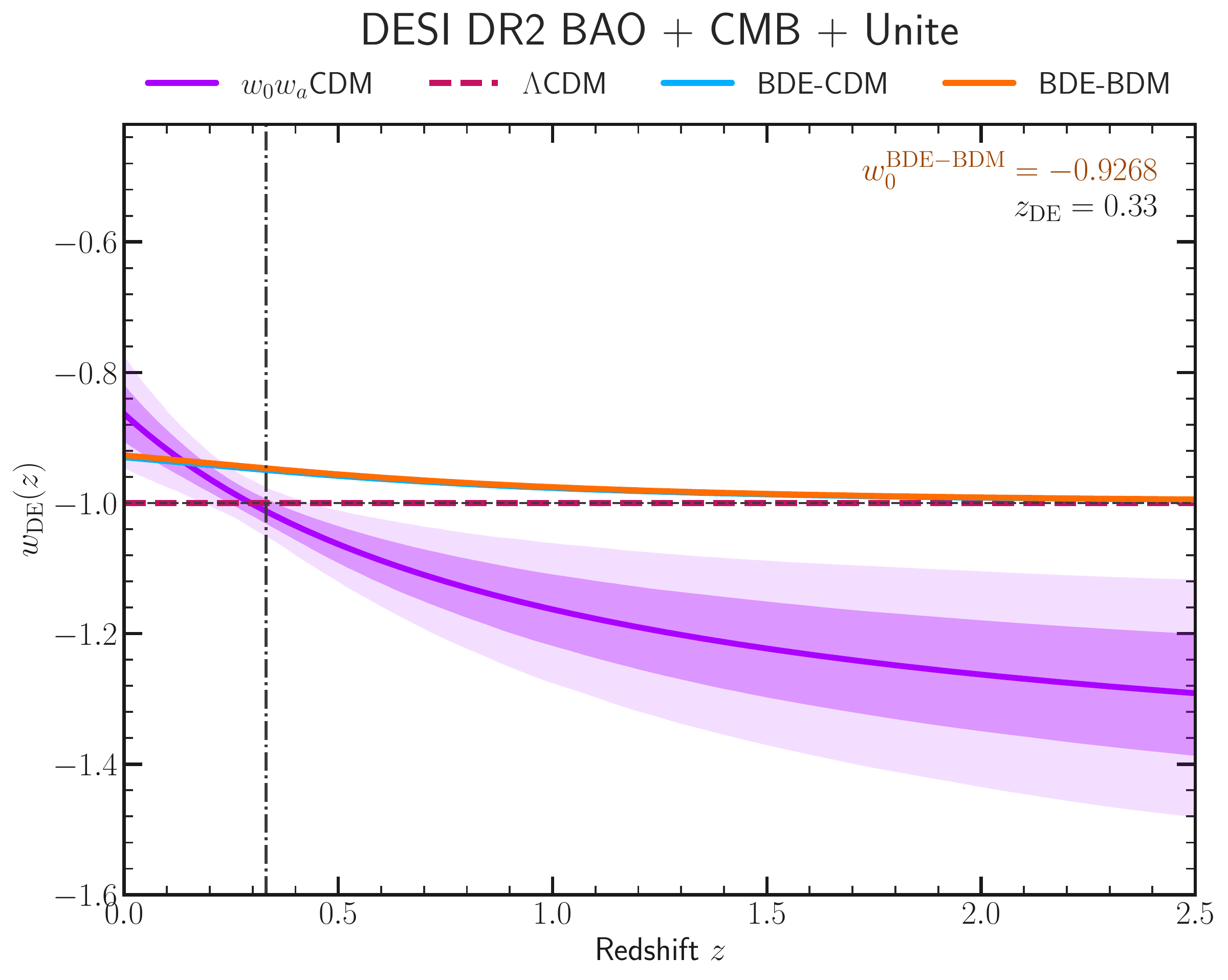}
\caption{\label{fig:wde}%
Dark-energy equation of state. (a) $w_{\rm DE}(a)$ from $a=10^{-6}$ to the present; the dotted line marks $\ac$ and
the dash-dotted line the onset of dark-energy domination. (b) $w_{\rm DE}(z)$ for $0\le z\le2.5$ with the $68\%$ and
$95\%$ bands; the BDE bands are narrower than the plotted lines.}
\end{figure}

Figure~\ref{fig:wde} shows the dark-energy equation of state. Over the full history [panel~(a)] the BDE models pass
through the stages described in section~\ref{sec:model}: radiation, $w=1/3$, until $\ac=2.50\times10^{-6}$; the
stiff phase, $w\simeq1$, until $a\simeq2\times10^{-3}$; and a frozen phase at $w\simeq-1$ through recombination and
most of the matter era. The \wowa{} fit, extrapolated to early times, lies in the phantom regime, $w\simeq-1.46$. At
late times [panel~(b)], the BDE-BDM equation of state rises monotonically to $w_0=-0.927$ today and remains above
$-1$, while the \wowa{} curve crosses $w=-1$ at $z=0.30$ and lies in the phantom regime at higher redshift; BDE-CDM and
BDE-BDM are indistinguishable at the scale of the figure. The two dynamical descriptions therefore agree only near
the present: the BDE equation of state is predicted with an uncertainty of a few $10^{-4}$, narrower than the plotted
line, whereas the \wowa{} band widens quickly with redshift. The dark energy dominates the expansion from
$z_{\rm DE}=0.33$ in BDE-BDM, against $0.32$ in \LCDM{} and $0.34$ in \wowa.

\begin{figure}[!tbp]
\centering
\includegraphics[width=0.325\textwidth]{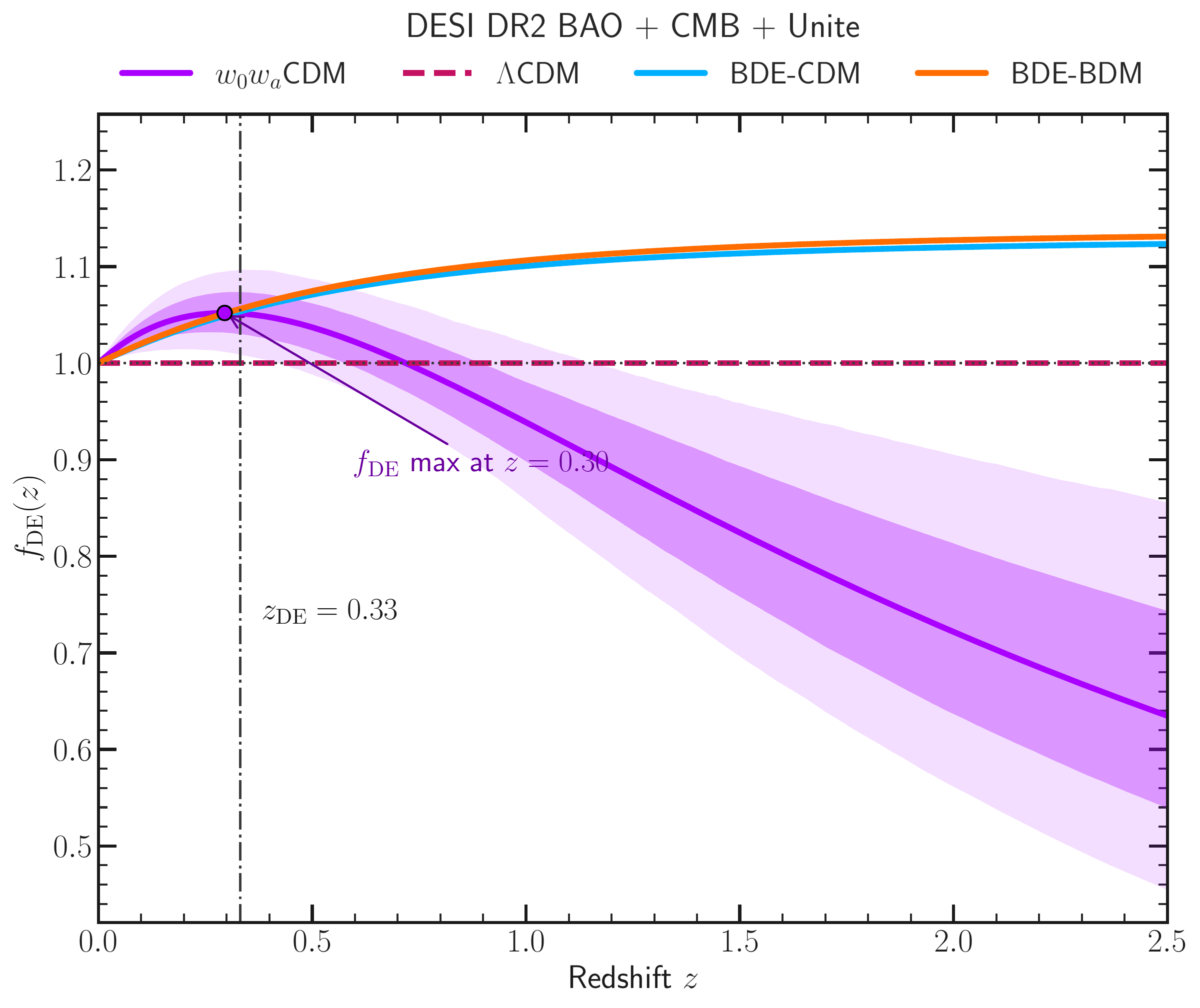}\hfill
\includegraphics[width=0.325\textwidth]{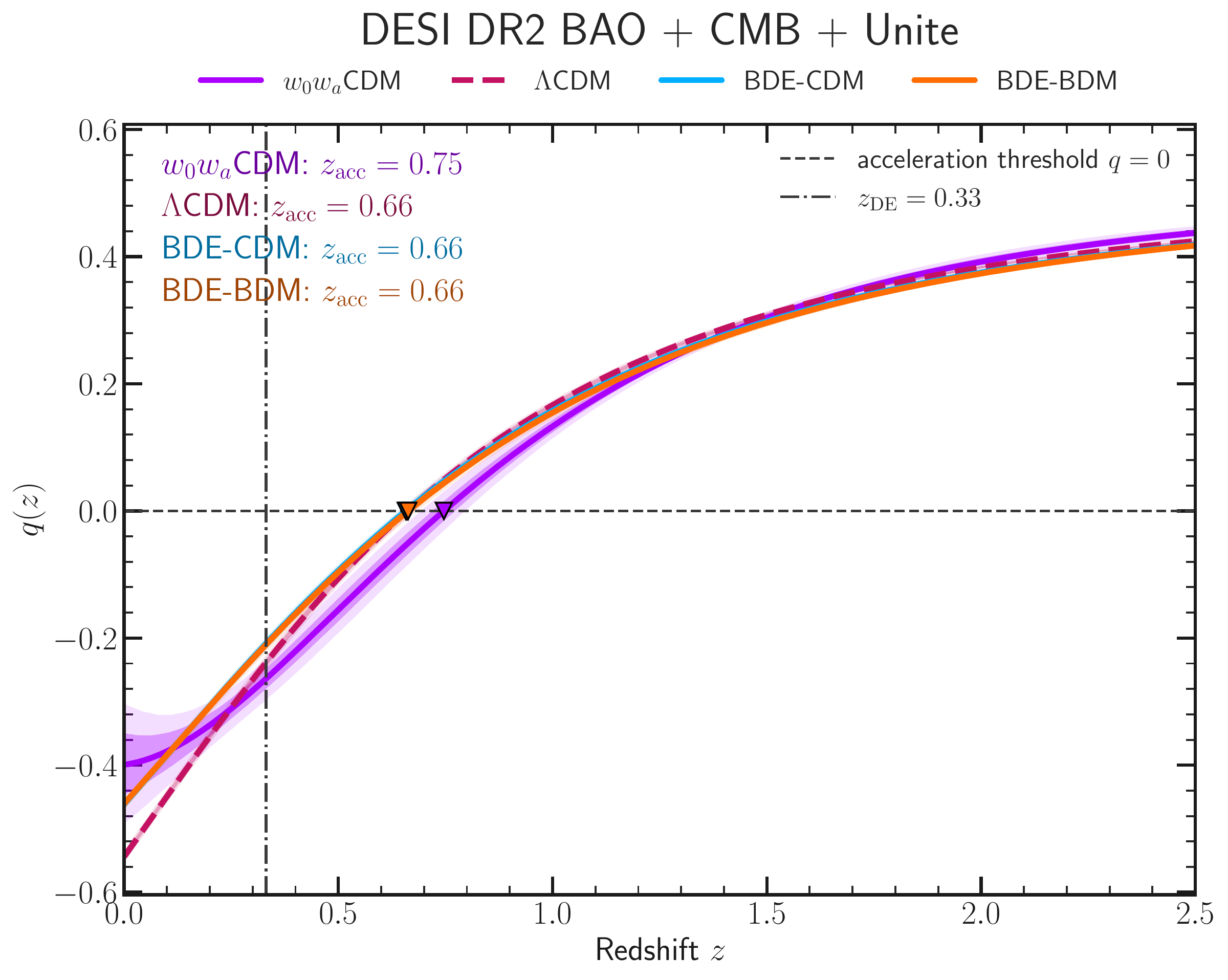}\hfill
\includegraphics[width=0.325\textwidth]{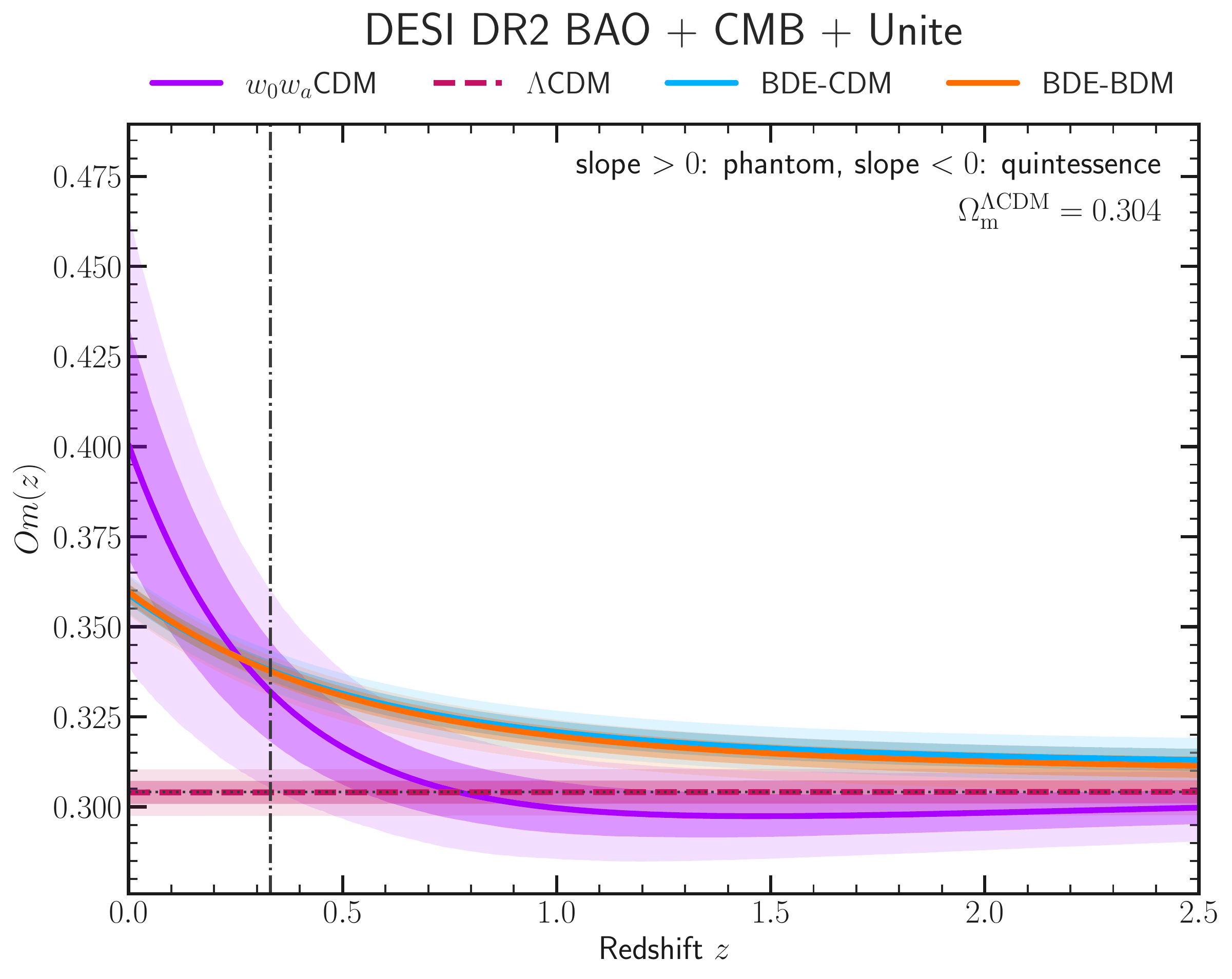}
\caption{\label{fig:expansion}%
(a) Dark-energy density relative to its present value, $f_{\rm DE}(z)$. (b) Deceleration parameter; the triangles
mark $z_{\rm acc}$. (c) The $Om(z)$ diagnostic. Bands are the $68\%$ and $95\%$ intervals.}
\end{figure}

Because $w>-1$ throughout, the BDE dark-energy density decreases monotonically with time, and
$f_{\rm DE}=\rho_{\rm DE}(z)/\rho_{\rm DE,0}$ grows with redshift, whereas in \wowa{} it peaks where the fitted
equation of state crosses $-1$ and decreases at higher redshift [figure~\ref{fig:expansion}(a)]. In BDE-BDM the dark energy
was $10\%$ denser at $z=1$ than today and $13\%$ denser at $z=2.5$, $f_{\rm DE}=1.10$ and $1.13$; in \wowa{} it peaks
at $f_{\rm DE}=1.05$ at $z=0.30$ and falls to $0.64$ at $z=2.5$. The BDE dark energy is therefore denser in the past than
today, whereas the \wowa{} dark energy is $36\%$ less dense at $z=2.5$, the redshift range of the DESI quasar and
Lyman-$\alpha$ BAO, where the \wowa{} reconstruction is least constrained. The deceleration parameter
[figure~\ref{fig:expansion}(b)] changes sign at $z_{\rm acc}=0.66$ in the BDE models, as in \LCDM, against $0.75$ in \wowa; the present
acceleration is weaker than in \LCDM, $q_0=-0.46$ against $-0.54$, and stronger than in \wowa, $-0.40$.
The $Om$ diagnostic~\cite{Sahni2008}, $Om(z)=[H^2(z)/\Hub^2-1]/[(1+z)^3-1]$, is constant for \LCDM{} and decreases with
redshift for $w>-1$; the BDE models show this negative slope at all
redshifts [figure~\ref{fig:expansion}(c)], falling from $Om=0.360$ today to $0.311$ at $z=2.5$, while \wowa{} falls more
steeply, from $0.400$ to $0.300$, and drops below the \LCDM{} value where its equation of state becomes phantom. The
slope of $Om(z)$ is a direct, model-independent test of $w>-1$, and its low-redshift value, set by the present
equation of state, is where the two dynamical models differ most.

\subsection{\texorpdfstring{Cosmological constraints in \LCDM, \wowa, BDE-CDM and BDE-BDM}{Cosmological constraints in LCDM, w0waCDM, BDE-CDM and BDE-BDM}}
\label{sec:constraints}

\begin{figure}[!t]
\centering
\includegraphics[width=0.58\textwidth]{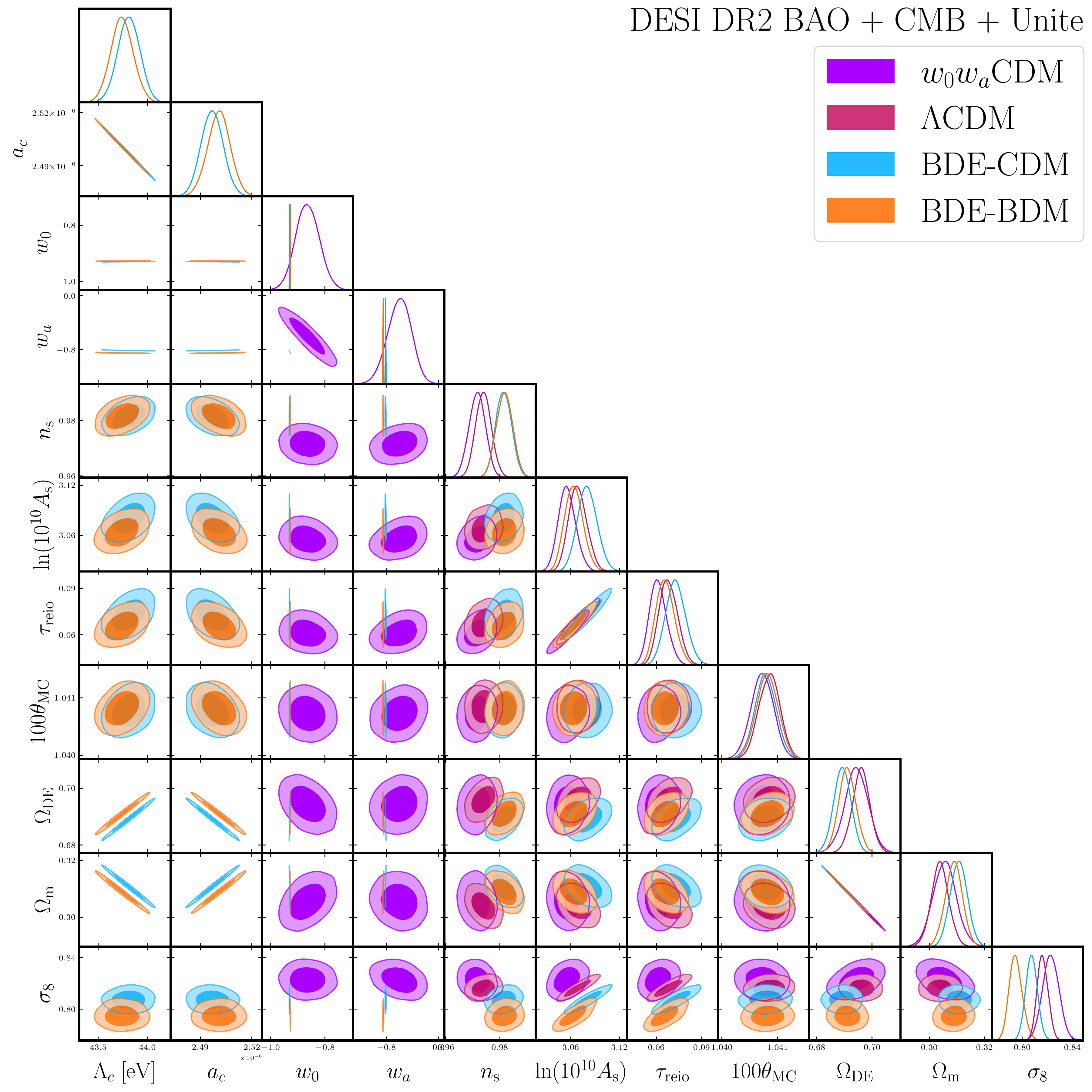}
\caption{\label{fig:triangle}%
Marginalized $68\%$ and $95\%$ credible regions for \wowa, \LCDM, BDE-CDM and BDE-BDM from DESI DR2 BAO $+$ CMB2026 $+$
Unite $+$ CMB lensing.}
\end{figure}

Table~\ref{tab:params} lists the constraints for the four models, and figure~\ref{fig:triangle} shows their joint
posteriors. The parameters shared with \LCDM{} move by less than $3\sigma$; the largest shift is that of the scalar
spectral index, $n_s=0.9817\pm0.0030$ against $0.9740\pm0.0029$ in \LCDM. It is shared by BDE-CDM, so it originates in
the dark-energy sector rather than in the bound dark matter. The amplitude
of structure is lowest in BDE-BDM, $\sig=0.795\pm0.005$ and $\Seight=0.807\pm0.007$, against $\sig=0.816\pm0.004$ and
$\Seight=0.822\pm0.006$ in \LCDM; about $40\%$ of the shift with respect to \LCDM{} is already present in BDE-CDM,
and the rest is due to the bound dark matter (section~\ref{sec:bdmcdm}). The \wowa{} posterior is the broadest, as
expected from its two additional parameters.

\begin{table}[!t]
\centering
\footnotesize
\setlength{\tabcolsep}{4pt}
\begin{tabular}{l cccc}
\hline
Parameter & $\Lambda$CDM & $w_0w_a$CDM & BDE-CDM & BDE-BDM \\
\hline
$\tau_{\rm reio}$ & $0.0681\pm0.0061$ & $0.0612\pm0.0057$ & $0.0729\pm0.0067$ & $0.0659\pm0.0061$ \\
$n_s$ & $0.9740\pm0.0029$ & $0.9716\pm0.0031$ & $0.9815\pm0.0029$ & $0.9817\pm0.0030$ \\
$\ln(10^{10}A_s)$ & $3.069\pm0.011$ & $3.056\pm0.011$ & $3.081\pm0.012$ & $3.064\pm0.011$ \\
\hline
$\Omega_m$ & $0.3039\pm0.0032$ & $0.3056\pm0.0042$ & $0.3107\pm0.0031$ & $0.3089\pm0.0031$ \\
$\sigma_8$ & $0.8163\pm0.0043$ & $0.8228\pm0.0065$ & $0.8075\pm0.0046$ & $0.7949\pm0.0053$ \\
$S_8$ & $0.8216\pm0.0062$ & $0.8304\pm0.0068$ & $0.8217\pm0.0062$ & $0.8066\pm0.0069$ \\
$w_0$ & $-1$ & $-0.8635\pm0.0433$ & $-0.9300\pm0.0003$ & $-0.9267\pm0.0003$ \\
$w_a$ & $0$ & $-0.599\pm0.181$ & $-0.811\pm0.003$ & $-0.847\pm0.004$ \\
\hline
$\Lambda_c$ [eV] & -- & -- & $43.81\pm0.11$ & $43.74\pm0.11$ \\
$10^{6}a_c$ & -- & -- & $2.4969\pm0.0064$ & $2.5009\pm0.0065$ \\
$f_{\rm BDE}$ & -- & -- & -- & $0.9398\pm0.0037$ \\
$v_c$ & -- & -- & -- & $0.37\pm0.14$ \\
\hline
\end{tabular}
\caption{\label{tab:params}%
Marginalized means and $68\%$ intervals for DESI DR2 BAO $+$ CMB2026 $+$ Unite $+$ CMB lensing. For the BDE models, $w_0$ and $w_a$ are the value and slope of $w(a)$ at $a=1$, eq.~\eqref{eq:w0waproj}; $a_c$ and $f_{\rm BDE}$ are derived. As in ref.~\cite{Camilleri2026}, we report only quantities that do not depend on an absolute distance scale for
this combination; $\Hub$, the physical densities $\omega_i=\Omega_ih^2$ and $r_{\rm d}$ are not quoted (section~\ref{sec:h0}).}
\end{table}

The comparison of the two BDE models in table~\ref{tab:params} shows how the bound dark matter acts on the parameters
shared with \LCDM. With cold dark matter, bound dark energy raises the optical depth to reionization and the primordial
amplitude, $\tau_{\rm reio}=0.0729\pm0.0067$ and $\ln(10^{10}A_s)=3.081\pm0.012$, against $0.0681\pm0.0061$ and
$3.069\pm0.011$ in \LCDM. With bound dark matter both return close to their \LCDM{} values,
$\tau_{\rm reio}=0.0659\pm0.0061$ and $\ln(10^{10}A_s)=3.064\pm0.011$, so that the unified model departs from \LCDM{}
only in the parameters that its dark sector is expected to modify: the expansion rate, the amplitude of structure and
the tilt.

The two dynamical descriptions of the DESI data also predict opposite shifts in the amplitude of structure. The
\wowa{} fit raises it above \LCDM, $\sig=0.823\pm0.007$ and $\Seight=0.830\pm0.007$, whereas BDE-BDM lowers it to $\Seight=0.806\pm0.006$, so
that the four models span $0.807\le\Seight\le0.830$ and BDE-BDM and \wowa{} differ by $2.5\sigma$ in $\Seight$.
Weak-lensing and cluster measurements of $\Seight$~\cite{descollaboration2026darkenergysurveyyear, refId0} are found to be consistent and no tension with value obtained from BDE-BDM model, serving as an independent consequence of the
dark-energy dynamics, complementary to the growth rate discussed in section~\ref{sec:growth}: in BDE-BDM both the
thawing dark energy and the free streaming of the bound dark matter act in the same direction, toward less
structure.

\FloatBarrier
\subsection{Distance--redshift and expansion rate measurements}
\label{sec:distances}

\begin{figure}[!tbp]
\centering
\includegraphics[width=0.47\textwidth]{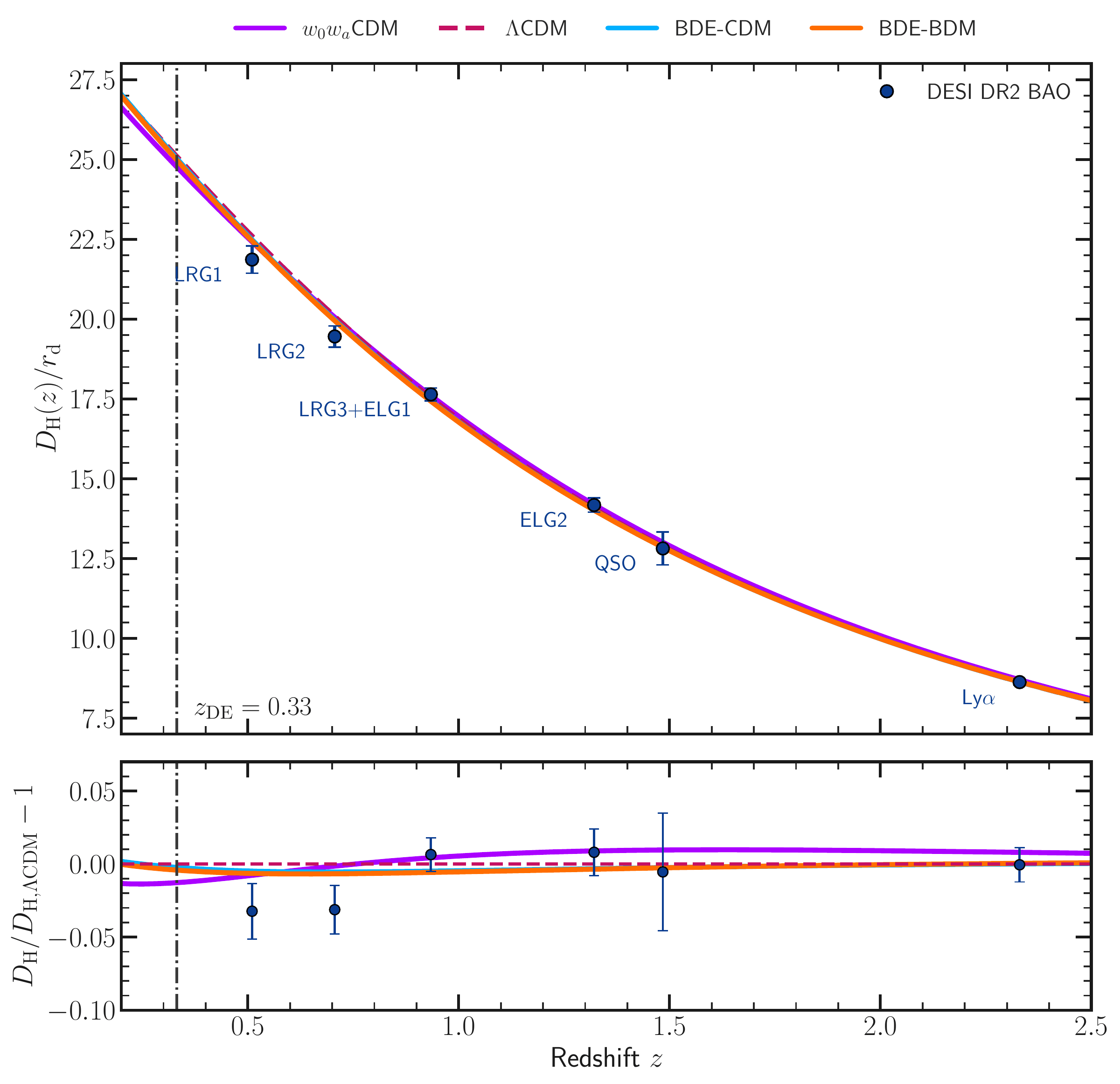}\hfill
\includegraphics[width=0.47\textwidth]{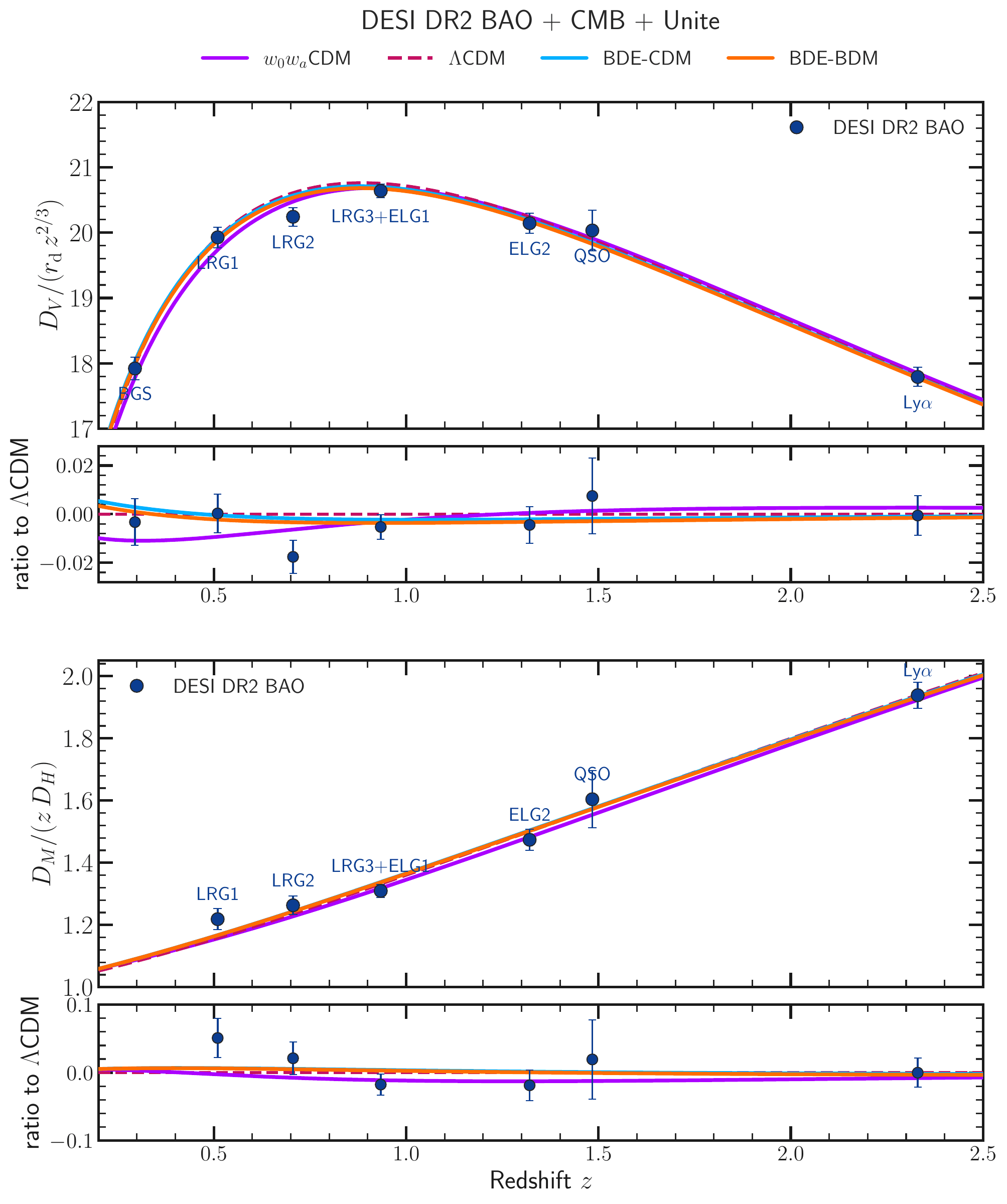}
\caption{\label{fig:hz}%
(a) Radial BAO distance $D_H(z)/r_{\rm d}$ against the DESI DR2 measurements~\cite{DESI2025II}, shown as measured; the curves are the
posterior means. (b) DESI DR2 distances $D_V/(r_{\rm d}z^{2/3})$ and $D_M/(zD_H)$. The lower panels show the ratios
to \LCDM.}
\end{figure}

Figure~\ref{fig:hz} compares the BAO distances of the four models with the DESI DR2 measurements, in the form in
which they are measured: the radial distance $D_H/r_{\rm d}=c/(Hr_{\rm d})$ in panel~(a), and the isotropic distance
$D_V/r_{\rm d}$ and the Alcock--Paczynski ratio $D_M/D_H$ in panel~(b). At the precision of the data the models are
close, and the differences appear in the ratios to \LCDM. Over the whole DESI range, the BDE-BDM distance ratios
$D_V/r_{\rm d}$, $D_M/r_{\rm d}$ and $D_H/r_{\rm d}$ stay within $0.9\%$, $1.0\%$ and $0.8\%$ of \LCDM.

\begin{figure}[!tbp]
\centering
\includegraphics[width=\textwidth]{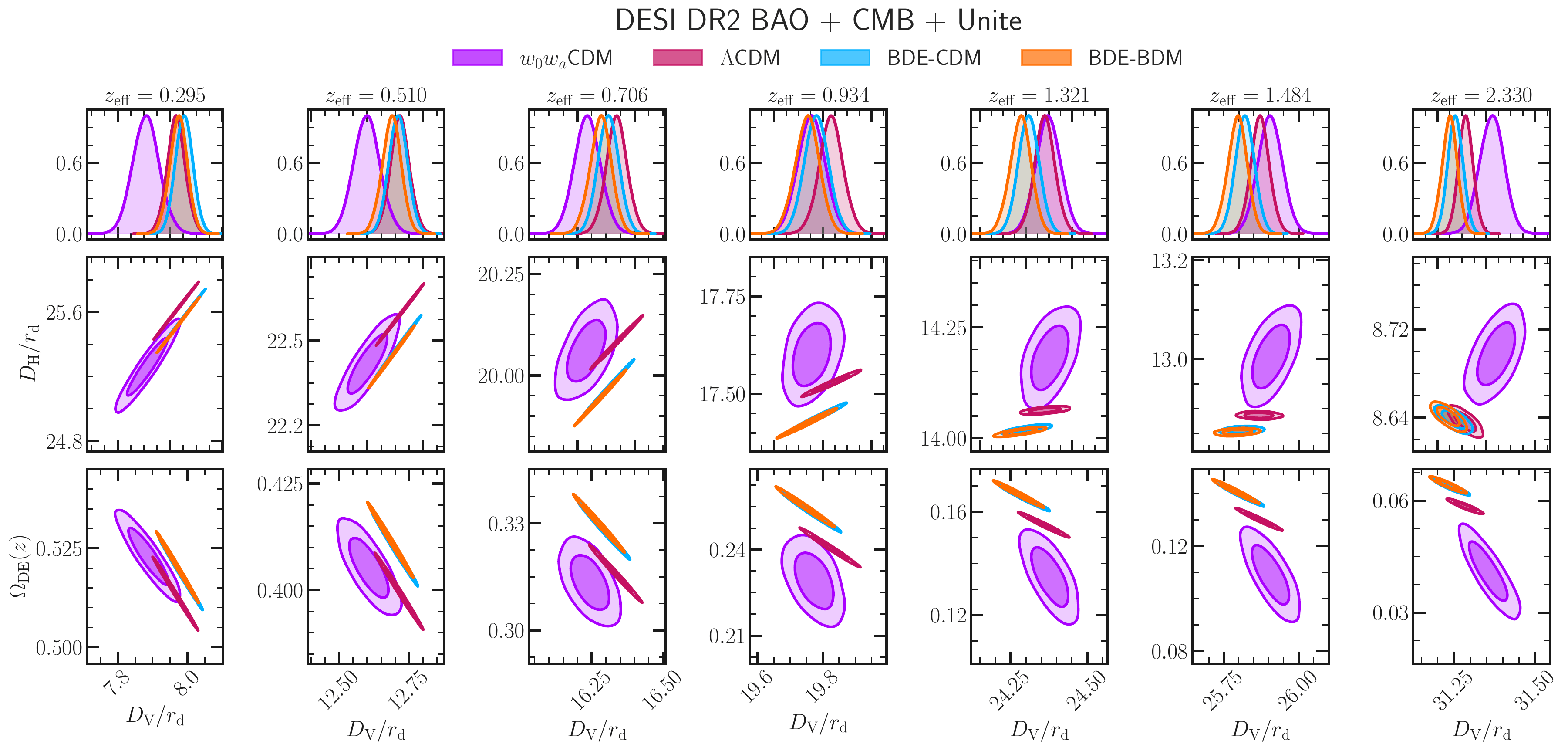}
\caption{\label{fig:dvgrid}%
Posterior predictions at the DESI DR2 effective redshifts: the distribution of $D_V/r_{\rm d}$, and its $68\%$
and $95\%$ regions with $D_H/r_{\rm d}$ and $\Omega_{\rm DE}(z)$.}
\end{figure}

These sub-percent differences are precisely what the DESI DR2 measurements resolve, and BDE-BDM gives the best fit
to them of the four models. At the joint best fits, the thirteen BAO measurements give $\chi^2_{\rm BAO}=10.97$ for
BDE-BDM, against $13.75$ for \LCDM, $11.51$ for \wowa{} and $13.22$ for BDE-CDM: a reduction of $2.78$, or $20.2\%$,
with respect to \LCDM, larger than the $16.3\%$ achieved by \wowa{} with two free parameters in the dark-energy
sector, and five times larger than that of BDE-CDM. The BAO preference for BDE-BDM is therefore not a consequence
of additional freedom: the equation of state is predicted by the gauge dynamics, and it is the combination of the
thawing dark energy with the bound dark matter that brings the predicted distances closest to the DESI
measurements. The posterior predictions at the DESI effective redshifts (figure~\ref{fig:dvgrid}) separate in the
quantities that $D_V$ averages over. BDE-BDM predicts $\Omega_{\rm DE}(z)$
higher than \LCDM{} by $0.006$--$0.014$ for $z\gtrsim0.5$ ($2.3\sigma$--$4.6\sigma$ of the combined posterior width),
the signature of a dark energy that was denser in the past, together with a lower radial distance; \wowa{} departs in
the opposite direction at high redshift, where its fitted equation of state is phantom.

The separation is concentrated in the radial direction. In the isotropic distance $D_V/r_{\rm d}$, the quantity
measured by the bright galaxy sample, BDE-BDM stays within $0.4\%$ of \LCDM{} at every DESI redshift, and \wowa{}
within $1.1\%$; the transverse distance $D_M/r_{\rm d}$ of BDE-BDM is within $0.31\%$ of \LCDM. The radial distance
$D_H/r_{\rm d}=c/(Hr_{\rm d})$, instead, separates the models in opposite directions: BDE-BDM lies $0.35\%$--$0.67\%$
below \LCDM{} for $0.5\lesssim z\lesssim1.5$, at $2.3\sigma$--$7.9\sigma$ of the combined posterior width, whereas
\wowa{} lies $0.4\%$--$1.0\%$ above it for $z\gtrsim0.9$. The radial BAO measurements of the luminous red galaxies,
emission-line galaxies and quasars are therefore the most direct discriminant between a thawing dark energy and the
phantom past of the \wowa{} fit.

\subsection{Matter power spectrum and structure formation}
\label{sec:growth}

\begin{figure}[!tbp]
\centering
\includegraphics[width=0.52\textwidth]{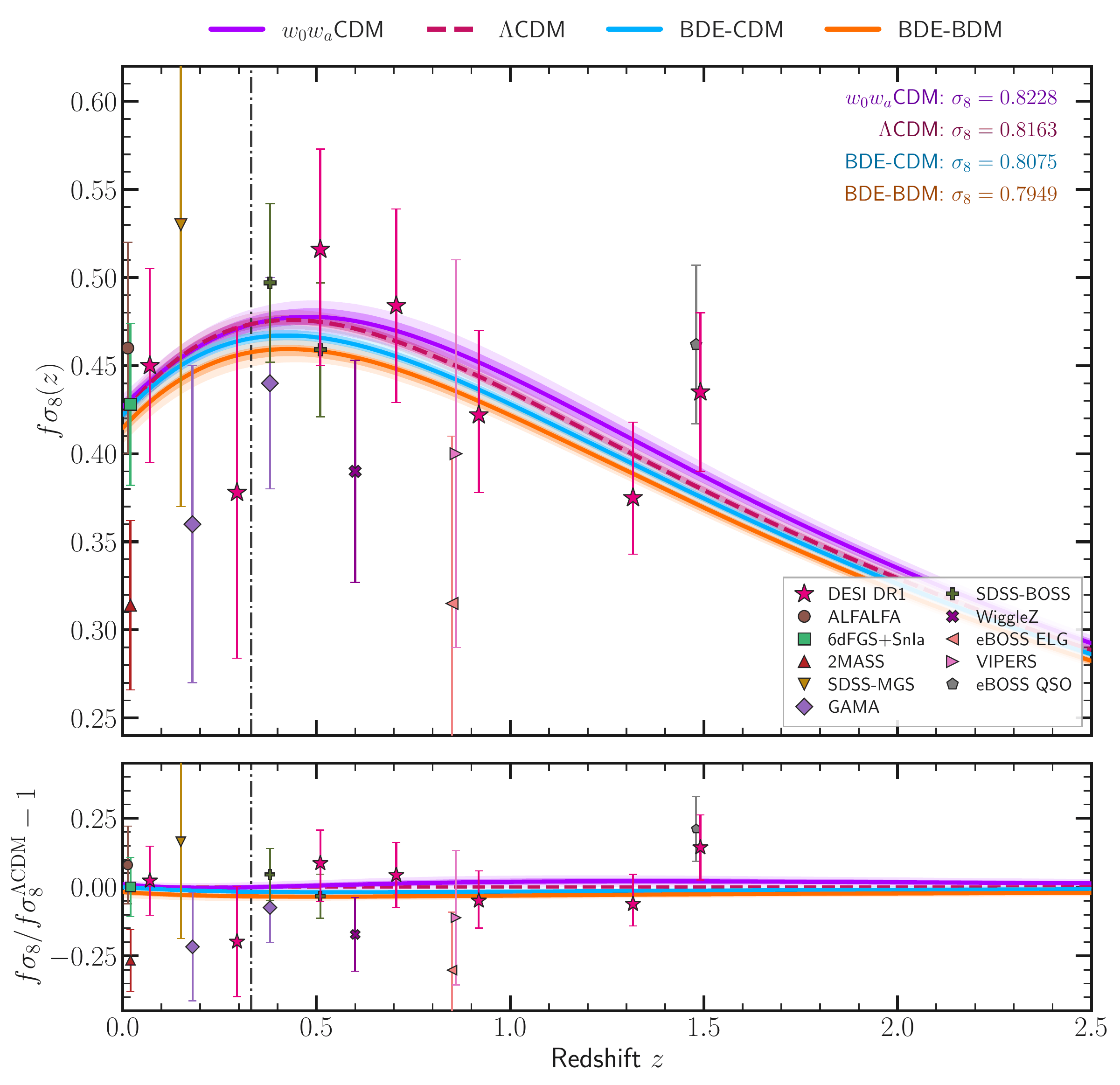}
\caption{\label{fig:fs8}%
Growth rate $f\sigma_8(z)$ against redshift-space distortion and peculiar-velocity measurements from
ALFALFA~\cite{Avila:2021ALFALFA}, 6dFGS and SNe~Ia~\cite{Huterer:2017SNvel}, 2MASS~\cite{Davis:2011}, the SDSS main
galaxy sample~\cite{Howlett:2015MGS}, GAMA~\cite{Blake:2013GAMA}, SDSS-BOSS~\cite{Alam:2017BOSS},
WiggleZ~\cite{Blake:2012WiggleZ}, the eBOSS emission-line galaxies~\cite{deMattia:2021eBOSS} and
quasars~\cite{Neveux:2020eBOSS}, VIPERS~\cite{Pezzotta:2017VIPERS} and DESI DR1~\cite{DESI:2024jxi,Lai:2026DESIPV},
with the fractional deviation from \LCDM{} in the lower panel.}
\end{figure}

\begin{figure}[!tbp]
\centering
\includegraphics[width=\textwidth]{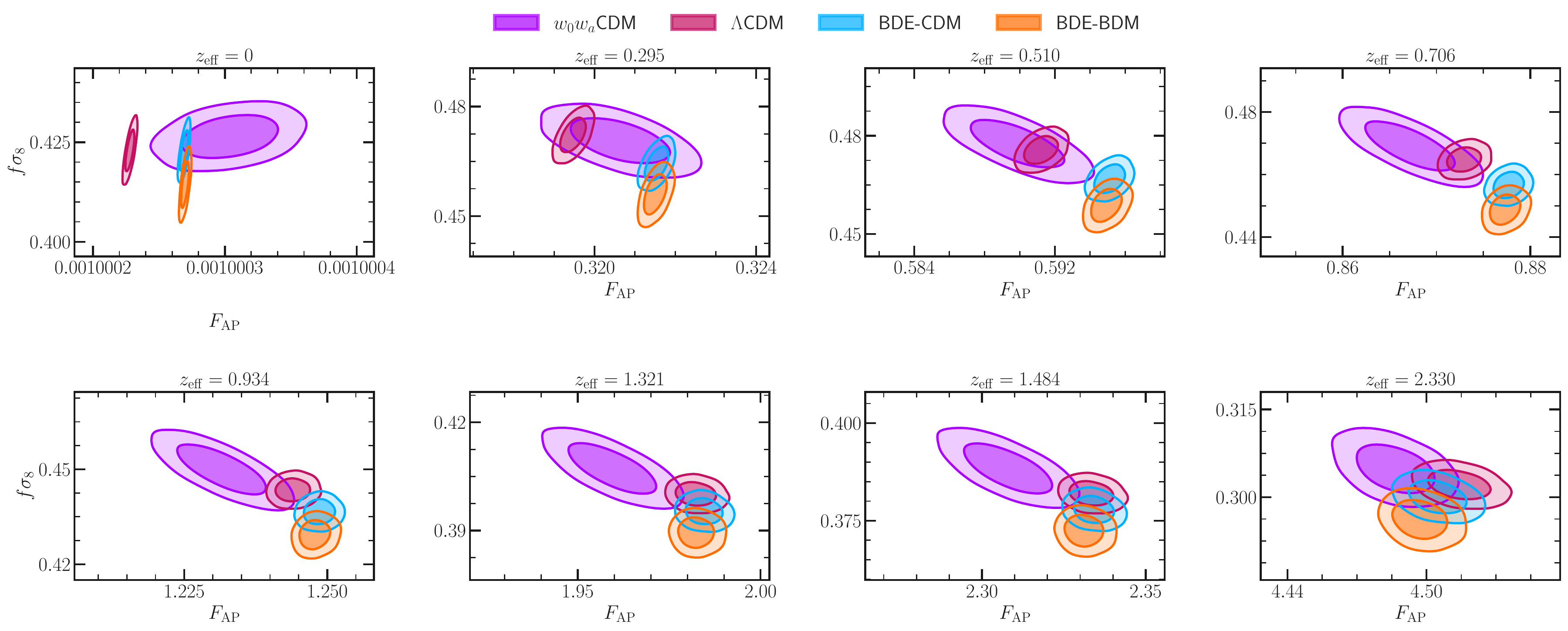}
\caption{\label{fig:fapgrid}%
Posterior predictions in the $F_{\rm AP}$--$f\sigma_8$ plane at $z=0$ and at the DESI DR2 effective redshifts.}
\end{figure}

The lower amplitude of structure translates into a lower growth rate (figure~\ref{fig:fs8}). At the DESI effective
redshifts (figure~\ref{fig:fapgrid}), BDE-BDM predicts $f\sigma_8$ below \LCDM{} by $0.007$--$0.017$, at
$2.6\sigma$--$4.1\sigma$ of the combined posterior width; BDE-CDM lies below \LCDM{} by about half as much
($1.0\sigma$--$2.3\sigma$), so the bound dark matter doubles the deficit. \wowa{} predicts $f\sigma_8$ within
$1.7\sigma$ of \LCDM, but a lower Alcock--Paczynski parameter $F_{\rm AP}=D_MH/c$ at $z\gtrsim0.7$, by
$1.9\sigma$--$2.8\sigma$, whereas the BDE models predict $F_{\rm AP}$ above \LCDM{} at low redshift and consistent with
it for $z\gtrsim0.9$. The growth rate peaks at
$z\simeq0.43$ in all models, at $f\sigma_8=0.459$ in BDE-BDM against $0.476$ in \LCDM; the redshift-space distortion
measurements in figure~\ref{fig:fs8} have uncertainties of $5\%$ or more, larger than this $3.5\%$ difference, so the
current data cannot yet discriminate. The two dynamical scenarios therefore depart from \LCDM{} along different directions in the plane
of geometry and growth, which a joint measurement such as the DESI full-shape analysis can separate.

The product $f\sigma_8$ separates two effects of the dark sector. The growth rate $f=d\ln D/d\ln a$ is lower than in
\LCDM{} by $0.3\%$--$1.2\%$ at the DESI redshifts, for instance $f=0.853$ against $0.862$ at $z=0.934$, and almost
identically in BDE-CDM and BDE-BDM: it is a consequence of the dark energy, which, being denser in the past, lowers
$\Omega_m(z)$ and with it the rate at which perturbations grow. The growth index defined by $f=\Omega_m(z)^\gamma$
remains close to its \LCDM{} value, $\gamma=0.5424$ against $0.5412$ at $z=0.510$, so the growth still follows
$\Omega_m(z)$ as in general relativity with smooth dark energy. The amplitude $\sigma_8(z)$, instead, is lower by
$2.0\%$--$2.6\%$ in BDE-BDM, at $2.2\sigma$--$3.1\sigma$ of the combined posterior width, but only by
$0.6\%$--$1.1\%$ in BDE-CDM: this part of the deficit is the free streaming of the bound dark matter. The two
components act in the same direction, which is why the predicted $f\sigma_8$ lies below \LCDM{} at every redshift.

\begin{figure}[!tbp]
\centering
\includegraphics[width=0.495\textwidth]{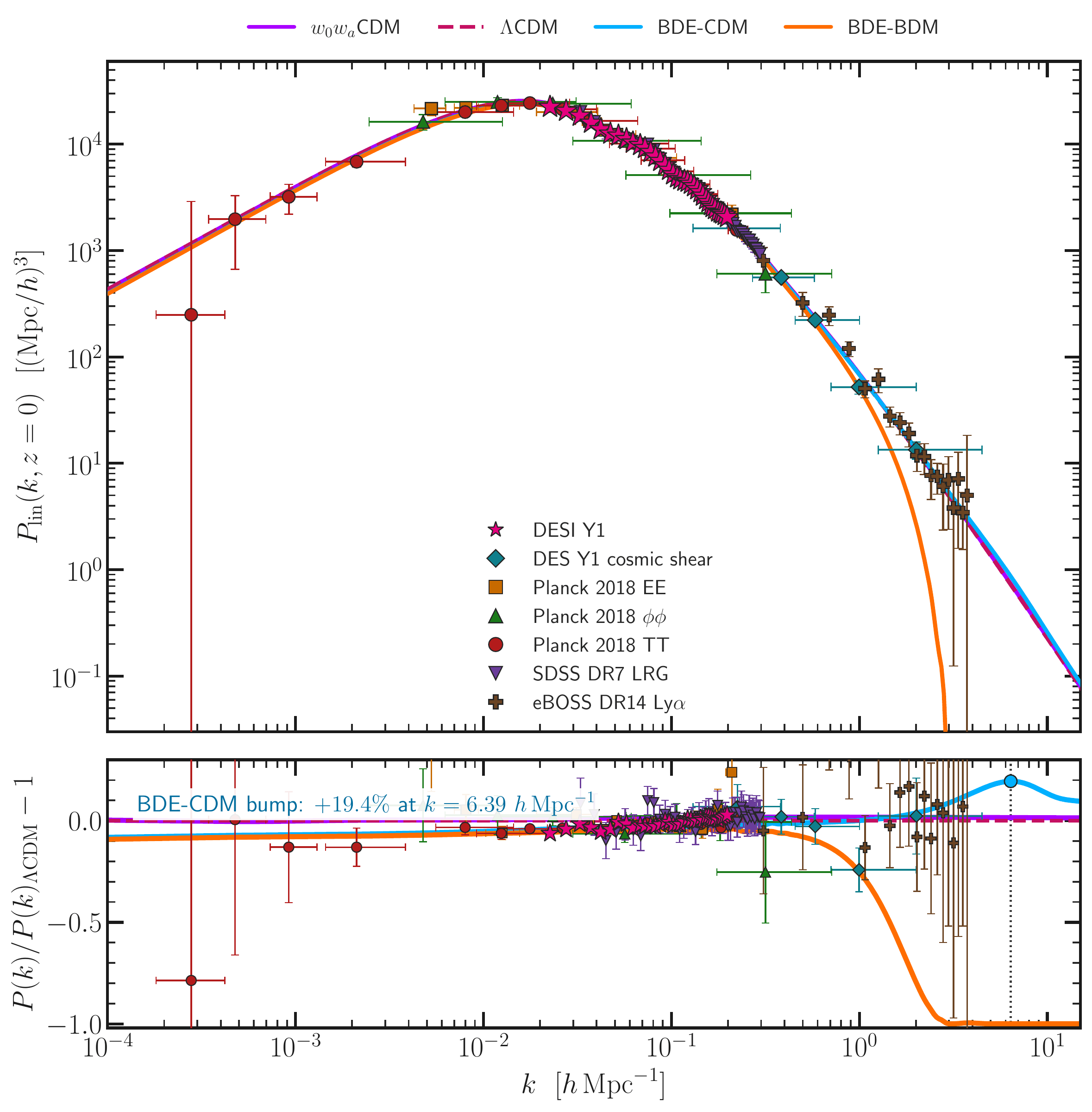}\hfill
\includegraphics[width=0.495\textwidth]{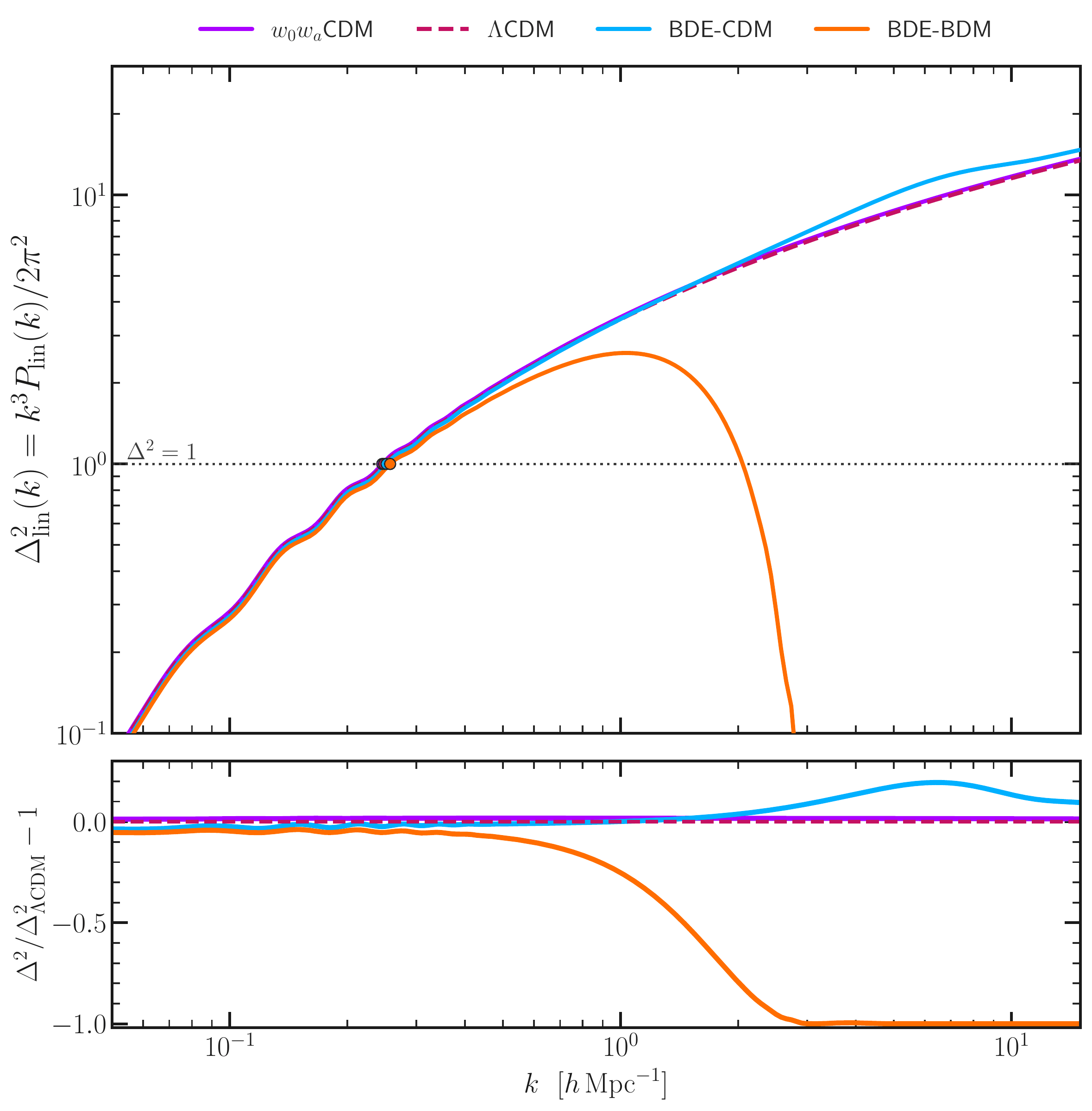}
\caption{\label{fig:pk}%
(a) Linear matter power spectrum at $z=0$ with the reconstructions from \textit{Planck} 2018~\cite{Planck2020V}, SDSS
DR7 LRG, DES Y1 cosmic shear and eBOSS DR14 Lyman-$\alpha$ data compiled in ref.~\cite{Chabanier2019}, and from DESI
Y1~\cite{Cereskaite_2026}. (b) Dimensionless spectrum $\Delta^2(k)=k^3P(k)/2\pi^2$. The lower
panels show the fractional deviation from \LCDM.}
\end{figure}

Figure~\ref{fig:pk} shows the linear matter power spectrum at $z=0$. On large scales the BDE models follow \LCDM{} up to
the lower amplitude. On small scales the two differ qualitatively. In BDE-CDM, the rapid dilution of the meson after
condensation enhances the growth of modes that entered the horizon before $\ac$, producing a bump of $+19.5\%$ at
$k=6.5\,h\,{\rm Mpc}^{-1}$~\cite{Almaraz2019,MacorraLozano2026JCAP}. In BDE-BDM, the free streaming of the bound dark
matter removes this bump and cuts the spectrum off: relative to its own cold limit, $\vc\to0$, the power falls to one
half at
\begin{equation}
  \khalf=1.49\,h\,{\rm Mpc}^{-1},\qquad M_{1/2}=\frac{4\pi}{3}\bar\rho_m\left(\frac{\pi}{\khalf}\right)^3
  =3.4\times10^{12}\,M_\odot\,h^{-1},
  \label{eq:khalf}
\end{equation}
for the posterior mean $\vc=0.37$, with $\bar\rho_m$ the present mean matter density; at $k=1\,h\,{\rm Mpc}^{-1}$ the
power is $77\%$ of the cold limit. With respect to \LCDM, the BDE-BDM spectrum is $4.5\%$ lower at
$k=0.1\,h\,{\rm Mpc}^{-1}$, $25\%$ lower at $k=1\,h\,{\rm Mpc}^{-1}$ and $79\%$ lower at $k=2\,h\,{\rm Mpc}^{-1}$,
while BDE-CDM is already $4\%$ higher at $k=2\,h\,{\rm Mpc}^{-1}$, on the rise to its bump. On the largest scales both
BDE spectra lie $5\%$--$8\%$ below \LCDM{} in units of $(h^{-1}{\rm Mpc})^3$, reflecting their lower
amplitude and the different value of $h$ in the units of the figure. The dimensionless spectrum [figure~\ref{fig:pk}(b)] shows that the cutoff lies well inside the nonlinear
regime: $\Delta^2=1$ at $k=0.257\,h\,{\rm Mpc}^{-1}$ in BDE-BDM, against $0.248\,h\,{\rm Mpc}^{-1}$ in \LCDM, a factor of
six below $\khalf$, so the free streaming affects galaxy-scale rather than BAO-scale clustering.

The figure also shows the linear power spectrum reconstructed from eBOSS DR14 Lyman-$\alpha$ data~\cite{Chabanier2019},
which extends to the scales of the BDE-BDM cutoff. These points are derived assuming a \LCDM{} shape and are not a
likelihood that can be applied to BDE-BDM, and no Lyman-$\alpha$ data are included in our analysis. A Lyman-$\alpha$
analysis carried out within the model is the decisive test of $\vc$, since the cutoff of eq.~\eqref{eq:khalf} falls
within the range of scales probed by the forest and such analyses are left for future work.

\FloatBarrier
\subsection{CMB temperature power spectrum}
\label{sec:cmb}

\begin{figure}[!hb]
\centering
\includegraphics[width=0.5\textwidth]{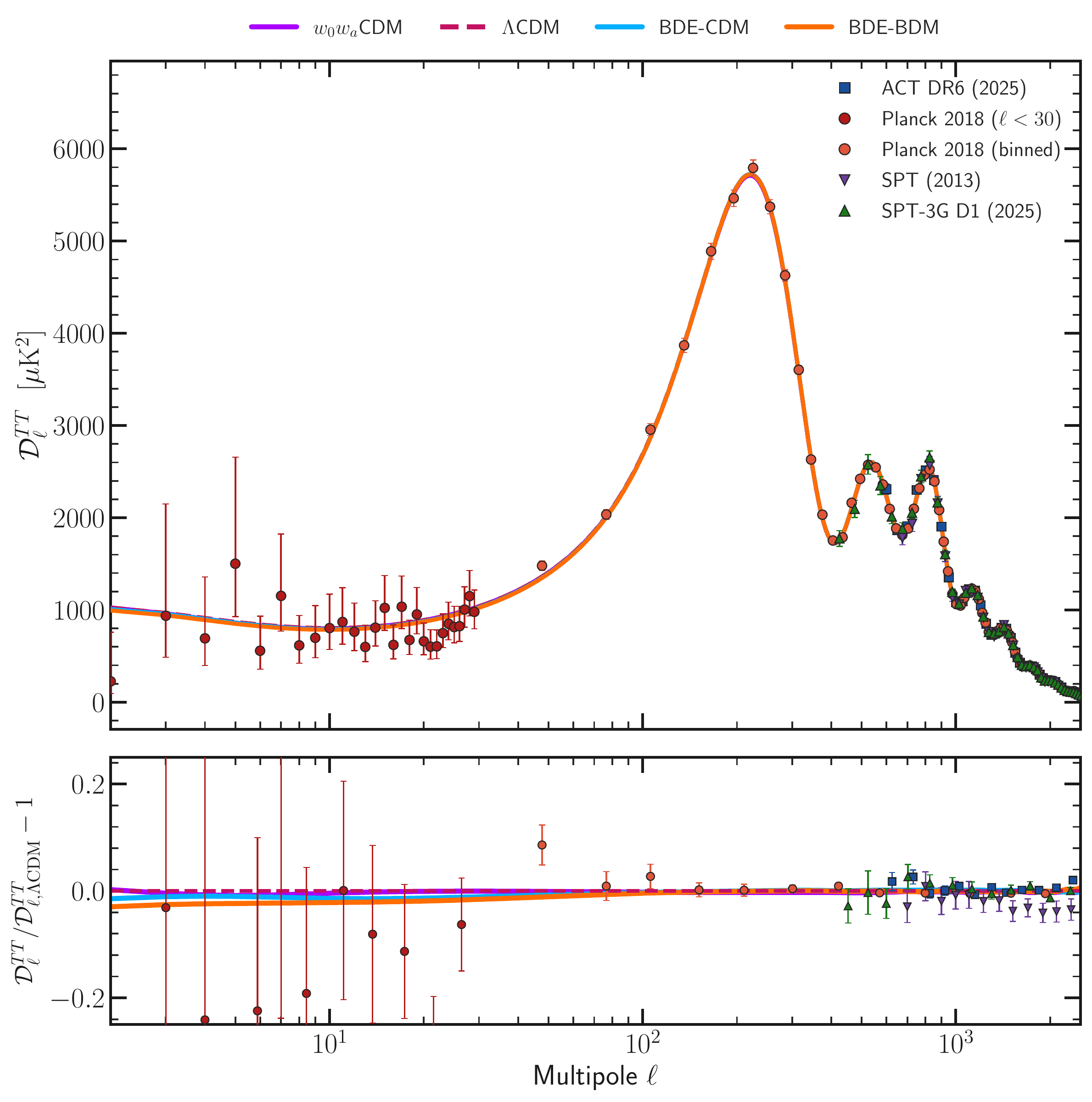}
\caption{\label{fig:cmbtt}%
Lensed CMB temperature spectrum against the \textit{Planck} 2018~\cite{Planck2020V}, ACT DR6~\cite{Louis2025},
SPT-3G D1~\cite{Camphuis2025} and SPT~\cite{Story:2012wx} measurements, with the fractional deviation from \LCDM{}
in the lower panel.}
\end{figure}

Figure~\ref{fig:cmbtt} compares the lensed CMB temperature spectra of the four models with the \textit{Planck}, ACT
and SPT measurements. At the precision of the figure the models are indistinguishable: BDE-BDM departs from \LCDM{} by
less than $3\%$ at $\ell<30$, well inside the cosmic variance, by less than $1.6\%$ across the acoustic peaks and by
less than $0.3\%$ in the damping tail. The angular scale of the sound horizon, which the positions of the peaks fix
to $0.02\%$, is the same in all models: $100\theta_{\rm MC}=1.04082\pm0.00020$ in BDE-BDM against $1.04085\pm0.00019$ in
\LCDM. This is expected from the history of the bound dark energy, which is negligible from shortly after
condensation until well after recombination (section~\ref{sec:eos}), so that the acoustic physics is that of \LCDM.
In the BDE models $\theta_{\rm MC}$ is not sampled but derived from $\Lc$ and the densities, so the precision with
which the acoustic scale is measured carries over to the constraint on the condensation scale.

The differences between the models are thus at the level of the measurement uncertainties, where the likelihoods
discriminate between them (table~\ref{tab:comparison}). With respect to \LCDM, BDE-BDM improves the fit to the
low-multipole temperature and polarization spectra, by $-0.95$ and $-1.47$ in $\chi^2$, and to the CMB lensing,
by $-0.43$, is indifferent to SPT-3G ($-0.02$), and pays a cost in the \textit{Planck} CamSpec and ACT DR6
high-multipole spectra, $+2.82$ and $+1.75$; the CMB as a whole costs $+1.7$ in $\chi^2$. For BDE-CDM the same cost
is $+8.8$, dominated by ACT DR6 ($+5.35$), CamSpec ($+3.28$) and the lensing ($+1.24$). Replacing cold by bound dark
matter therefore lowers the CMB penalty of bound dark energy by about $7$, through ACT DR6, the lensing and the
low-$\ell$ polarization; this is the CMB part of the preference for bound dark matter discussed in
section~\ref{sec:comparison}. \wowa{} fits the CMB slightly better than \LCDM, by $-6.2$, with improvements spread over
most of the likelihoods. The remaining cost of the BDE models, common to both, is concentrated in the \textit{Planck}
CamSpec spectra; the two models also share a high spectral index, $n_s=0.982$ against $0.974$ in \LCDM{}
(section~\ref{sec:constraints}).

\section{Model comparison}
\label{sec:comparison}

Table~\ref{tab:comparison} lists the best-fit $\chi^2$ of each likelihood, the DIC and the Bayes factors. Three features
stand out.

\begin{table}[!t]
\centering
\small
\begin{tabular}{l r rrr}
\hline
 & $\Lambda$CDM & \multicolumn{3}{c}{$\Delta$ relative to $\Lambda$CDM} \\
Likelihood & $\chi^2$ & $w_0w_a$CDM & BDE-CDM & BDE-BDM \\
\hline
DESI DR2 BAO & 13.75 & $-2.24$ & $-0.53$ & $-2.78$ \\
Unite SNe & 2783.49 & $-4.80$ & $+0.16$ & $+0.48$ \\
Planck low-$\ell$ TT & 21.97 & $+0.20$ & $-0.79$ & $-0.95$ \\
Planck low-$\ell$ EE & 392.38 & $-1.71$ & $-0.18$ & $-1.47$ \\
Planck CamSpec & 3289.27 & $-0.49$ & $+3.28$ & $+2.82$ \\
ACT DR6 & 142.04 & $-1.60$ & $+5.35$ & $+1.75$ \\
SPT-3G D1 & 162.23 & $-1.27$ & $-0.11$ & $-0.02$ \\
CMB lensing & 38.74 & $-1.29$ & $+1.24$ & $-0.43$ \\
\hline
Total $\chi^2_{\rm min}$ & 6843.87 & $-13.20$ & $+8.41$ & $-0.60$ \\
$\Delta$DIC & -- & $-8.78$ & $+8.96$ & $+1.46$ \\
$\ln B$ vs $\Lambda$CDM & -- & $-0.36$ & $-3.69$ & $-0.50$ \\
\hline
\end{tabular}
\caption{\label{tab:comparison}%
Best-fit $\chi^2$ per likelihood for $\Lambda$CDM and differences for the other models, with the deviance information criterion and the Bayes factor against $\Lambda$CDM ($H_0\sim\mathcal{U}[20,100]$ in place of $100\theta_{\rm MC}$; section~\ref{sec:methods}). Negative $\Delta\chi^2$ and $\Delta$DIC and positive $\ln B$ favor the model over $\Lambda$CDM. In addition, $\ln B_{\rm BDE\text{-}BDM,\,BDE\text{-}CDM}=+3.19$ and $\ln B_{\rm BDE\text{-}BDM,\,w_0w_aCDM}=-0.15$.}
\end{table}

First, BDE-BDM provides the best fit to the DESI DR2 BAO of all four models, $\Delta\chi^2_{\rm BAO}=-2.78$ with
respect to \LCDM, a $20.2\%$ improvement, against $-2.24$ for \wowa{} and $-0.53$ for BDE-CDM
(section~\ref{sec:distances}). Beyond the BAO, it improves or matches the \LCDM{} fit in five of the eight
likelihoods: the low-$\ell$ temperature and polarization spectra ($-0.95$ and $-1.47$), the CMB lensing ($-0.43$)
and SPT-3G ($-0.02$). The cost is concentrated in the \textit{Planck} CamSpec and ACT DR6 high-$\ell$ spectra
($+2.82$ and $+1.75$) and, marginally, in the supernovae ($+0.48$), and the total is $\Delta\chi^2=-0.60$.

Second, BDE-BDM is statistically equivalent to both \LCDM{} and \wowa{}, while being far more predictive than the
latter. Its deviance information criterion differs from that of \LCDM{} by $\Delta{\rm DIC}=+1.46$, below the
threshold of $2$ at which differences become meaningful, and its Bayes factors are $\ln B=-0.50$ against \LCDM{} and
$-0.15$ against \wowa, both well inside the inconclusive range $|\ln B|<1$; with the physically motivated prior
$\vc\le1/\sqrt2$ they become $-0.16$ and $+0.19$. This equivalence is achieved with no free parameter in the
dark-energy sector, where \wowa{} has two: the large $\chi^2$ improvement of \wowa, $-13.2$, of which $-4.8$ comes
from the Unite supernovae, does not translate into evidence, $\ln B=-0.36$, because it is paid for by its two
additional parameters. The data therefore do not prefer the unconstrained parametrization over a model in which the
dark-energy history is predicted from the gauge dynamics.

Third, bound dark matter is clearly preferred over cold dark matter. With the same bound dark energy, replacing cold
by bound dark matter lowers $\chi^2$ by $9.0$ and gives $\ln B_{\rm BDE\text{-}BDM,\,BDE\text{-}CDM}=+3.19$ ($+3.53$
with $\vc\le1/\sqrt2$), moderate evidence on the scale of ref.~\cite{Trotta2008}. The improvement is broad: about $7$
of the $9.0$ come from the CMB, where the ACT DR6 penalty falls from $+5.4$ to $+1.8$, the lensing term from $+1.2$ to
$-0.4$ and the low-$\ell$ EE term improves by $1.3$, and the remaining $2.3$ from the BAO, whose gain over \LCDM{}
grows from $-0.5$ to $-2.8$. Bound dark energy with cold dark matter is disfavored against \LCDM, $\Delta\chi^2=+8.4$
and $\ln B=-3.69$, in agreement with ref.~\cite{LozanoMacorra2026MNRAS}; it is the unification of the dark energy and
the dark matter in a single condensation that makes the bound dark energy competitive with \LCDM{} on these data.

Compared with our analysis of DESI DR2 with \textit{Planck} and three earlier supernova
compilations~\cite{LozanoMacorra2026PRD}, the BAO preference for BDE-BDM persists with the new data; the overall
balance has shifted toward statistical equivalence with \LCDM{} because the new high-multipole CMB spectra and the
Unite supernovae are less favorable to the BDE models than the data used there.

\section{Conclusions}
\label{sec:conclusions}

We have confronted the unified bound dark energy--bound dark matter model with the most recent cosmological data:
the DESI DR2 baryon acoustic oscillations, the \textit{Planck}, ACT DR6 and SPT-3G D1 CMB spectra, the combined ACT
and SPT lensing reconstruction and the Unite supernova compilation. In this model the dark energy and the dark matter
originate in the condensation of a single supersymmetric SU(3) dark gauge group: the lightest dark meson drives the
late-time acceleration, and the heavier bound states, created with a velocity $\vc$, form the dark matter. Because
the condensation scale is fixed by gauge-coupling unification, the dark-energy sector has no free parameter beyond
those of \LCDM, and the dark-matter sector adds only $\vc$.

The data confirm the central prediction of the model. The measured condensation scale, $\Lc=43.74\pm0.11$~eV, lies
inside the window $34^{+16}_{-11}$~eV predicted by unification, and the condensation leaves $94\%$ of its energy in
the dark energy, $\fbde=0.9398\pm0.0037$. The resulting dark energy is of the thawing, non-phantom kind: frozen at
$w\simeq-1$ through most of cosmic history, it rises to $w_0=-0.9267\pm0.0003$ today and never crosses the phantom
divide. It reproduces the DESI distances through a history that flattens toward $-1$ at high redshift, where the
\wowa{} parametrization requires a phantom past, and its dark-energy density was $13\%$ higher at $z=2.5$ than today. The model does not, however, address the
Hubble tension: from the BAO and CMB alone, $\Hub=67.22\pm0.22$~km\,s$^{-1}$\,Mpc$^{-1}$, $5.5\sigma$ below the
local distance-ladder value, since the dark energy leaves the sound horizon unchanged.

The model provides the best description of the DESI DR2 baryon acoustic oscillations among the four models
considered, reducing $\chi^2_{\rm BAO}$ by $20.2\%$ with respect to \LCDM, more than \wowa{} with its two free
parameters. Over the full data combination it is statistically equivalent to both \LCDM{} and \wowa{} in $\chi^2$,
DIC and Bayesian evidence, while predicting its dark-energy history rather than fitting it. Within bound dark energy,
the bound dark matter is moderately preferred over cold dark matter, $\ln B=+3.2$, a preference that comes mostly from
the CMB and that makes the unified model, unlike bound dark energy with cold dark matter, competitive with \LCDM.

The new small-scale CMB and lensing data constrain the bound velocity from both sides, $\vc=0.37\pm0.14$, excluding
both the cold limit and the warm, semi-relativistic regime at $95\%$. The corresponding free-streaming cutoff of the
matter power spectrum lies at $\khalf=1.49\,h\,{\rm Mpc}^{-1}$, and the amplitude of structure is lower than in
\LCDM, $\Seight=0.807\pm0.007$. As a consequence, the model predicts a growth rate $f\sigma_8$ below \LCDM{} by
$2.6\sigma$--$4.1\sigma$ of the current posterior width at the DESI effective redshifts, with a geometry close to that
of \LCDM, the opposite pattern to \wowa, which departs in the Alcock--Paczynski parameter while keeping the growth
rate of \LCDM.

These predictions make the unified dark sector testable in the near future, where upcoming Lyman-$\alpha$
forest analyses and small-scale weak lensing will probe the free-streaming cutoff that fixes the bound velocity.
Together they will test whether a single dark gauge group can account for both the dark energy and the dark matter
of the Universe.

\acknowledgments

We thank the Unite team for sharing their data prior to publication. We acknowledge financial support from PAPIIT-DGAPA-UNAM (101124), SECIHTI project CBF-2025-G-1299, and the SECIHTI
scholarship grant for Jose Lozano.

\paragraph{Data availability.} The data that support the findings of this article are openly available and referenced
in the text.

\bibliographystyle{JHEP}

\providecommand{\href}[2]{#2}\begingroup\raggedright\endgroup

\end{document}